\documentclass[twocolumn,times]{aastex63}
\usepackage[utf8]{inputenc}
\usepackage[normalem]{ulem}
\usepackage{amsmath}

\newcommand{\HII}{H\,{\sc ii}\ }

\begin{document}

\title{Implications for Galactic Electron Density Structure from Pulsar Sightlines Intersecting \HII Regions}

\author[0000-0002-4941-5333]{Stella Koch Ocker}
\affiliation{Cahill Center for Astronomy and Astrophysics, California Institute of Technology, Pasadena, CA 91125, USA}
\affiliation{Carnegie Science Observatories, Pasadena, CA 91101, USA}

\author[0000-0001-8800-1793]{Loren D. Anderson}
\affiliation{Department of Physics and Astronomy, West Virginia University, Morgantown, WV 26506, USA}

\author{T. Joseph W. Lazio}
\affiliation{Jet Propulsion Laboratory, California Institute of Technology, Pasadena, CA 91106, USA}

\author[0000-0002-4049-1882]{James M. Cordes}
\affiliation{Department of Astronomy and Cornell Center for Astrophysics and Planetary Science, Cornell University, Ithaca, NY, 14853, USA}

\author[0000-0002-7252-5485]{Vikram Ravi}
\affiliation{Cahill Center for Astronomy and Astrophysics, California Institute of Technology, Pasadena, CA 91125, USA}
\affiliation{Owens Valley Radio Observatory, California Institute of Technology, Big Pine, CA 93513, USA}

\correspondingauthor{Stella Koch Ocker}
\email{socker@caltech.edu, socker@carnegiescience.edu}

\begin{abstract}
    Recent radio surveys have revealed pulsars with dispersion and scattering delays induced by ionized gas that are larger than the rest of the observed pulsar population, in some cases with {electron column densities (or dispersion measures; DMs)} larger than the maximum predictions of Galactic electron density models. By cross-matching the observed pulsar population against \HII region catalogs, we show that the majority of pulsars with $\rm DM > 600$ pc cm$^{-3}$ and scattering delays $\tau(1\ {\rm GHz}) > 10$ ms lie behind \HII regions, and that \HII region intersections may be relevant to as much as a third of the observed pulsar population. {The fraction of the full pulsar population with sightlines intersecting \HII regions is likely larger.} Accounting for \HII regions resolves apparent discrepancies where Galactic electron density models place high-DM pulsars beyond the Galactic disk. By comparing emission measures (EMs) inferred from recombination line observations to pulsar DMs, we show that \HII regions can contribute tens to hundreds of pc cm$^{-3}$ in electron column density along a pulsar LOS. We find that nearly all pulsars with significant excess (and deficit) scattering from the mean $\tau$-DM relation are spatially coincident with known discrete ionized gas structures, including \HII regions. Accounting for \HII regions is critical to the interpretation of radio dispersion and scattering measurements as electron density tracers, both in the Milky Way and in other galaxies. 
\end{abstract}

\keywords{ISM:general --pulsars --radio transients --\HII regions --scattering --turbulence}

\section{Introduction}\label{sec:intro}

The Milky Way interstellar medium (ISM) contains a range of ionized structures, from highly dense \HII regions enshrouding ultraviolet stars and tenuous hot bubbles evacuated by supernovae (SNe), to the diffuse, warm ionized medium that extends both across and beyond the Galactic plane. The distribution of this multi-phase gas is a critical tracer of star formation and stellar feedback, and their role in sculpting our Galaxy. Radio pulsars are sensitive probes of ionized gas, which causes dispersion, scattering, and birefringence manifesting as frequency-dependent time delays and intensity modulations in observed radio pulses. Accounting for these propagation effects is critical to pulsar timing experiments \citep{ng15_noise_budget,epta_noise_budget,ppta_noise_budget}, in addition to serving as input to Galactic electron density models that predict distances to both Galactic and extragalactic radio sources \citep{cordes91,ne20011,ymw16}.

Recently, high-sensitivity pulsar surveys with MeerKAT and the Five-hundred-meter Aperture Spherical radio Telescope (FAST) have unveiled pulsars in the Galactic plane with dispersion and scattering delays larger than the rest of the known population \citep{johnston2020,oswald2021_meerkat,han2021_fast,posselt2023_meerkat}. In some cases, the pulse dispersion measure (${\rm DM} = \int_0^D n_e dl$, where $n_e$ is electron density and $D$ is the pulsar distance) is larger than the maximum DMs predicted by Galactic electron density models NE2001 \citep{ne20011,ne20012} and YMW16 \citep{ymw16}, implying that these pulsars either lie beyond the Galactic disk or that the models are missing a significant portion of the ISM's free electrons along these lines-of-sight (LOSs). The FAST Galactic Plane Pulsar Survey (GPPS) has thus far identified $\sim 10$ pulsars with DMs larger than Galactic model predictions that are spatially coincident with \HII regions {in the Cygnus complex}, suggesting that these \HII regions may explain the apparent DM excess \citep{han2021_fast}. 

\HII regions can dramatically influence pulsar propagation observables by enhancing dispersion, scattering, and Faraday rotation, which are all LOS-integrated effects. Based on their DMs and rotation measures (RMs), pulsars have been used to infer mean electron densities $\approx 10$ cm$^{-3}$ \citep{ocker2020} and magnetic field magnitudes $\approx 6$ $\mu$G \citep{harveysmith2011} within \HII regions. Localization of pulsar scattering to \HII regions suggests that density fluctuations within those regions are moderately enhanced relative to the warm ionized ISM \citep{ocker2020,mall2022,ocker2024_scintarcs} and may be consistent with a turbulence cascade that dissipates at an inner scale $\approx 70 - 100$ km \citep{rickett2009}. Unambiguously making such inferences for individual \HII regions requires knowledge of both the pulsar and \HII region distances, and is hindered by the limited pulsar distance sample and the large kinematic uncertainties of many \HII region distances. Constraints on ionized gas elsewhere along the pulsar LOS are also generally required, based on examination of other pulsars at similar locations and/or analysis of complementary multi-wavelength surveys sensitive to diffuse ionized gas. As such, pulsars have been used to probe the internal properties of \HII regions in only a select handful of cases \citep[e.g.][]{mitra2003,harveysmith2011,ocker2020}, even though \HII regions have long been invoked as a potential explanation for enhanced DM and scattering in the inner Galaxy \citep{spangler1991,mckee1997}. 

In this study, we leverage expansive \HII region catalogs to perform a large-scale comparison between the Galactic distribution of \HII regions and the pulsar population. Such a comparison is now fruitful due largely to the $> 2\times$ increase in the number of both known and likely \HII regions over the past decade, chiefly thanks to infrared (IR) and radio surveys sensitive to \HII regions across the entire Galactic plane \citep{bania2010,wisecat_2014,wenger2021} and a concurrent increase in the known Galactic pulsar population. We employ two large \HII region catalogs, the WISE Catalog \citep{wisecat_2014}, {so called for its use of the Wide-Field Infrared Survey Explorer \citep{wright2010}}, and \cite{houhan2014}, hereafter HH14, which together contain $\approx 8500$ \HII regions identified at IR, radio, and/or optical wavelengths, of which $\approx 3000$ have measured distances. Despite uncertainties in both pulsar and \HII region distances, we are able to identify hundreds of pulsars that likely intersect \HII regions, of which $\sim 100$ pulsars have both DM and scattering measurements. We assess the impact of these \HII regions on pulsar observations through both a statistical analysis of the sample, and through detailed investigation of specific regions. 

The paper is organized as follows: Section~\ref{sec:catalogs} introduces the \HII region catalogs and their distance measurements. Section~\ref{sec:intersections} describes the methodology used to identify pulsars intersecting these \HII regions and describes the number and validity of intersections. In Section~\ref{sec:trends} we discuss general trends in the radio properties of pulsars intersecting \HII regions, and in Section~\ref{sec:gdigs_fast} we compare pulsar properties to radio recombination line observations of specific regions based on the Green Bank Diffuse Ionized Gas Survey (GDIGS; \citealt{anderson2021_gdigs}), FAST GPPS \citep{hou2022_fast}, and the Sino-German 6 cm Survey (SGS; \citealt{gao2019}). Section~\ref{sec:scattering} uses pulsar scattering measurements to constrain the strength of electron density fluctuations within \HII regions. {Implications for Galactic electron density models and their applications are discussed in Section~\ref{sec:implications}, and conclusions are summarized in Section~\ref{sec:summary}.}

\section{Galactic \HII Region Catalogs}\label{sec:catalogs}

While \HII regions were first identified in large numbers by their optical emission \citep{gum1955,sharpless1959}, modern catalogs have leveraged IR and radio surveys to discover thousands of \HII regions that are not detected at optical wavelengths due to the severe visual extinction at their relatively large distances. We employ two comprehensive \HII region catalogs, the WISE Catalog {(v2.3, \citealt{wisecat_2014}}) and HH14 \citep{houhan2014}, which encompass \HII region detections across optical, IR, and radio wavelengths. A subset of these \HII regions have distance measurements described further below, and which form the primary sample that we compare to pulsar LOSs. {Sources in the WISE Catalog were selected from IR data, spanning all Galactic longitudes and latitudes $|b| \leq 8^\circ$.  The sources were cross-matched with known \HII regions and in the process known \HII regions at higher (and lower) absolute latitudes were added.}
By contrast, HH14 is a compilation of \HII region detections from the published literature, based on surveys spanning a range of wavelengths, sensitivities, and sky coverages. The overlap between \HII regions with distances cataloged in WISE and HH14 is incomplete, in part because HH14 includes spectrophotometric distances whereas WISE does not. In order to make use of as many \HII region distances as possible, we use both catalogs and account for their overlap in subsequent analyses (\S\ref{sec:intersections}). Many of the kinematic distances we use originate from sources adopting different Galactic rotation curves. While re-deriving kinematic distances for the entire \HII region sample using a single Galactic rotation curve is beyond the scope of this work, we note that the differences between kinematic distances derived from different rotation curve parameters are typically comparable to or smaller than most kinematic distance uncertainties \citep[e.g.][]{houhan2014}. Each catalog is described in turn below. 
{HH14 is available on the Vizier catalog service \citep{vizier:J/A+A/569/A125}, and we use v2.3 of the WISE Catalog available at \url{http://astro.phys.wvu.edu/wise/} with distances computed by \cite{wenger_database}.}  %and is available on the Vizier catalog service \citep{vizier}; {for the latest version of the WISE Catalog, see \url{http://astro.phys.wvu.edu/wise/}}.

\subsection{WISE Catalog}

The WISE Catalog consists of $\approx 8400$ \HII regions detected using {IR data from the WISE and Spitzer space telescopes.} WISE {and Spitzer} can detect \HII regions across the entire Galaxy due to a combination of low extinction and high sensitivity. {Centroid positions and angular sizes for the \HII regions are provided by the catalog based on the IR data (in contrast to HH14, which only provides centroids).} The catalog is estimated to be complete for all \HII regions ionized by O-stars, and likely has a high completeness fraction for stellar types B0-B2 \citep{wisecat_2014,armentrout2021}. The mid-IR emission is from heated dust and exhibits a characteristic morphology consisting of {$\sim20~\mu$m} emission surrounded by {$\sim10~\mu$m} emission that coincides with radio continuum emission \citep{anderson2012}. In practice, \HII region morphologies are complex and varied (see e.g. \citealt{wood1989}), and they can be confused with planetary nebulae and external galaxies if their spectral energy distributions are not taken into account. The WISE Catalog excludes such interlopers where possible based on cross-matches with both known objects and surveys of recombination lines, which are unambiguous identifiers of \HII regions. About $2000$ \HII regions in the WISE Catalog {are classified as ``known'' based on} cross-matching with the \cite{fich1990} H$\alpha$ emission catalog covering the Northern hemisphere and via radio recombination line (RRL) observations \citep{caswell1987,lockman1989,lockman1996,bania2010,bania2012,anderson2015,alves2015,anderson2018,liu2019,wenger2021}. These recombination line observations imply that $<5\%$ of sources  in the WISE Catalog {classified as ``candidates'' (corresponding to positive detection of IR and radio continuum emission)} are not bona fide \HII regions \citep{anderson2011}, {and that $>50\%$ of sources classified in the WISE Catalog as ``radio-quiet'' (only IR detections to date) are bona fide \HII regions \citep{armentrout2021}.} 

{\cite{wenger_database} provides $\approx 2800$ distances for \HII regions in the WISE Catalog,} with nearly 400 of those distances based on parallax and the rest kinematic. The most accurate distance determinations are from maser parallaxes, many provided by \cite{reid2019}. Kinematic distances include those based on RRL radial velocities measured by the \HII Region Discovery Survey \citep{wenger2021}. Kinematic distances have substantial uncertainties, with inner Galaxy LOSs yielding both near and far possible solutions. \cite{wisecat_2014} break the kinematic distance ambiguity (KDA) where possible using emission and absorption of HI and H$_2$CO \citep[e.g.][]{dunham2011,urquhart2011,urquhart2012}, {and if the \HII region is detected optically, e.g. in H$\alpha$ emission, the near distance is adopted}. Even when the KDA is broken, kinematic distances can still have uncertainties of up to a few kiloparsecs. For cases where the KDA is not resolved, we adopt the near distance as a lower limit and the far distance as an upper limit. We ignore near kinematic distances $<0.1$ kpc, which lead to many spurious pulsar intersections. 

\subsection{HH14 Catalog}

The HH14 Catalog provides $4550$ spectral line measurements towards $2540$ \HII regions, of which about $1870$ have distance measurements. HH14 is a compilation of primarily optical and radio detections from the literature, including a large fraction of RRLs that were also included in {the WISE Catalog}. The catalog does not include most of the IR-only candidates identified with WISE, but it does provide spectrophotometric distances in addition to kinematic and parallax distances, thereby increasing the total sample of \HII region distances at our disposal. {HH14 also includes distances for several large, high-latitude \HII regions without distances in the WISE Catalog (e.g. S7, S27).} Spectrophotometric distances are based on identification of the main ionizing star and its spectral type, with associated uncertainties arising from a combination of {misidentification of the ionizing star,} spectral (mis)-classification, photometric errors, spread in the stellar luminosity of a given spectral type, and the assumed reddening law \citep[e.g.][]{foster2015}. The impact of these distance uncertainties on our analysis is considered in Section~\ref{sec:candidates}. All sources in HH14 have ionized gas line measurements, and we thus expect nearly all sources in HH14 are bona fide \HII regions.

\section{Pulsars Intersecting \HII Regions}\label{sec:intersections}

We first describe the methodology used to identify \HII region intersections, followed by a summary of the intersection candidates identified and our assessment of their validity.

\subsection{Methodology}\label{sec:methods}

We identify pulsars intersecting \HII regions using three key parameters: the pulsar and \HII region sky positions, distances, and the \HII region size. We compare the pulsar population to the WISE and HH14 catalogs independently, and then merge the results to produce a single unique list of intersections.

\begin{figure}
    \centering
    \includegraphics[width=0.45\textwidth]{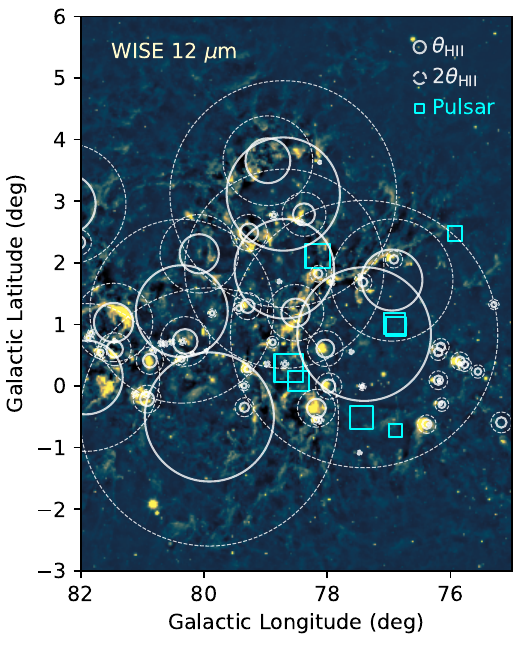}
    \caption{Illustration of the identification of pulsars likely behind \HII regions.  The background is a WISE 12\,$\mu$m image.  White circles are for \HII regions for which there are estimated distances, with the sizes of the circles based on their $\theta_{\mathrm{IR}}$ in the WISE Catalog.  Large white circles that appear to be devoid of 12\,$\mu$m emission are brighter at 20\,$\mu$m and identified as \HII regions on that basis. Cyan squares indicate pulsars identified as lying behind an \HII region based on the criteria described in Section~\ref{sec:methods}.  The sizes of the squares indicate the relative magnitudes of the pulsar DMs, with larger squares corresponding to larger DMs.}
    \label{fig:intersection_example}
\end{figure}

For the WISE Catalog cross-match, we exclude pulsars with projected offsets $>2 \theta_{\rm IR}$ away from the \HII region centroid position, where $\theta_{\rm IR}$ is the angular radius of the IR emission provided by the WISE Catalog. The maximum projected offset (i.e., the maximum impact parameter) is larger than the IR angular size because the exact physical extent of the ionized gas is not known a priori, and the IR angular size may be an underestimate of the ionized gas extent {due to radiation leakage that ionizes the outer envelope of the \HII region \citep[e.g.][]{luisi2019,dey2024}.} HH14 does not provide the \HII region angular sizes, so we set the maximum impact parameter to $20$ pc, re-scaled to an angular size based on the \HII region distance. This projected offset is the minimum value that reproduces known \HII region intersections from \cite{mitra2003}, \cite{harveysmith2011}, and \cite{ocker2020}, {which are cases where the pulsar and \HII region distances are known and the pulsar location is spatially coincident with the \HII region in H$\alpha$ emission maps.} An impact parameter cutoff of $20$ pc corresponds roughly to the expected {Str\"{o}mgren radii} of \HII regions around star types O8-O9 and B0, suggesting that our analysis may miss intersections through \HII regions around early-type O-stars \citep{rubin1968}. The incompleteness fraction of our intersection analysis is assessed further in Section~\ref{sec:candidates}. Figure~\ref{fig:intersection_example} shows an example of intersections through an \HII region complex in the WISE Catalog, identified using our selection criteria. 

We only search for intersections through the $\approx 3000$ \HII regions with distance measurements, but we consider all radio pulsars in the ATNF Catalog \citep[v2.1.1,][]{psrcat}\footnote{\url{http://www.atnf.csiro.au/research/pulsar/psrcat}}. Only about 300 pulsars have distance measures, half of which are parallax\footnote{\url{https://hosting.astro.cornell.edu/research/parallax/}} \citep[e.g.,][]{deller2019,ding2023} and the rest based on either H~{\sc i} kinematics or associations \citep[e.g.,][]{frail1990,pan2021_globs}. For pulsars without distance measures, we use Python implementations of NE2001 \citep{ne20011,ne20012,ocker2024_ne2001p} and YMW16 \citep{ymw16,price2021} to estimate their distances based on observed DMs, and we {classify the pulsar LOS as intersecting an \HII region if} at least one of the models predicts a pulsar distance greater than the \HII region distance, with the following caveats: For the sample of pulsars with known distances, these Galactic electron density models typically misestimate distances by $\approx30\%$ \citep{price2021}, although for some pulsar LOSs the models misestimate the distance by $\gtrsim100\%$ due to discrete, unmodeled density structures, including \HII regions \citep{ocker2020,ding2023}. {While NE2001 models \HII regions for select LOSs on an ad hoc basis, YMW16 does not model \HII regions at all.} When a pulsar does intersect an \HII region {that is not included in the models}, they will generally overestimate the pulsar distance because they have to integrate to larger distances to reproduce the observed DM. In these cases, the uncertainty in the model distance does not impact our ability to identify \HII region intersections. {Nonetheless, the combination of uncertainties on both pulsar and \HII region distances implies that we expect some false intersection candidates and some intersections that are missed altogether, and we attempt to identify these false positives and negatives where possible (\S\ref{sec:candidates}).} {Future work could constrain intersection probability distributions using distance priors \citep[e.g.][]{jennings2018,wenger2018}, which would obviate the need for false positive and negative rates.}

The intersection analysis is performed for two pulsar samples, the entire known radio pulsar population ($\sim 3000$ pulsars), and a subset of $\approx 500$ pulsars with scattering measurements. The first, bulk pulsar sample is used to statistically compare the DM distribution of intersection candidates against that of the broader pulsar population. The second sample of pulsars with scattering measurements is used for detailed analysis, where the sample is small enough to individually assess the validity of each intersection candidate based on the transverse proximity of the pulsar LOS to the \HII region and the difference between the pulsar's estimated distance and that of the \HII region (see below). 

\subsection{Intersection Candidates}\label{sec:candidates}

\begin{figure}
    \centering
    \includegraphics[width=0.48\textwidth]{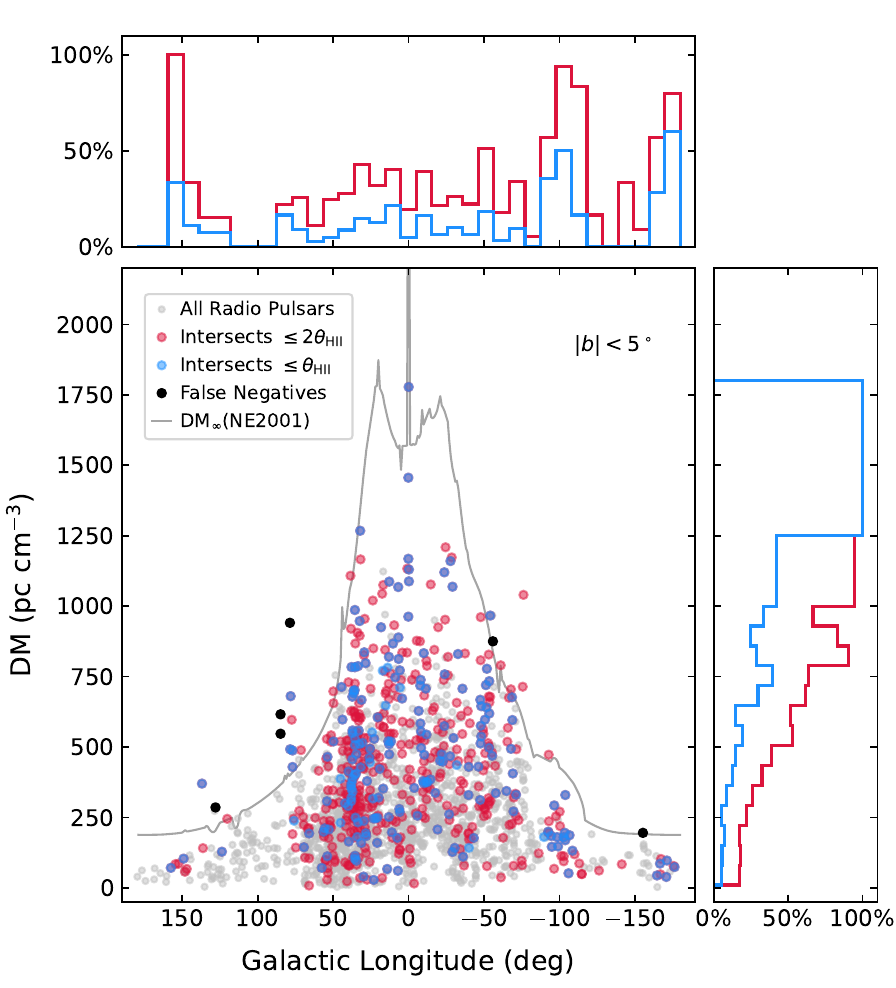}
    \caption{DM vs. Galactic longitude for all radio pulsars at Galactic latitudes $|b|<5^\circ$ (grey points) and for pulsars intersecting \HII regions with measured distances. Red points indicate pulsars that pass within $2\theta_{\rm HII}$ of a known \HII region, where $\theta_{\rm HII}$ is the angular radius provided by WISE, assumed to be 10 pc for \HII regions in HH14 ({corresponding to a nominal impact parameter cutoff $2\theta_{\rm HII} = 20$ pc}; \S\ref{sec:intersections}). Blue points are pulsars within $\theta_{\rm HII}$. The thin grey line indicates the maximum DM predicted by NE2001 at $b = 0^\circ$. {Black points indicate pulsars with large DMs that do intersect \HII regions but were missed by our selection criteria (false negatives).} The top and side panels show the fraction of all low-latitude pulsars that intersect \HII regions within $2\theta_{\rm HII}$ (red lines) and $\theta_{\rm HII}$ (blue lines).}
    \label{fig:dm_vs_lon}
\end{figure}

The initial cross-match between all known radio pulsars and the \HII regions with distances in the WISE and HH14 catalogs yields {630} intersection candidates, where intersections are taken to be impact parameters $<2\theta_{\rm IR}$ for \HII regions in the WISE Catalog or 20 pc for \HII regions in HH14. About $220$ ($33\%$) of these intersections are within half the impact parameter cutoff ($\theta_{\rm IR}$ for the WISE Catalog or 10 pc for HH14), but this more restrictive cutoff excludes a number of pulsars that intersect \HII region complexes with extended emission beyond the IR angular radii recorded in the WISE Catalog. Figure~\ref{fig:dm_vs_lon} shows the distribution of DM vs. Galactic longitude for the \HII region intersections identified with both impact parameter cutoffs; this figure is discussed further in Section~\ref{sec:trends}. We estimate a false positive rate of $\approx 20\%$ for spurious intersections (see below).

We identify $106$ intersections where the pulsar has a published scattering measurement. We build upon the compilation of scattering measurements from \cite{cordes2016,cordes2022}, which encompasses $\approx 500$ pulsars that have either pulse broadening times and/or scintillation bandwidths reported in the literature, by adding scattering times we have inferred from pulse profiles released by FAST GPPS \citep{han2021_fast}. The list of \HII region intersections is shown in Table~\ref{tab:intersections1}, including both pulsar and \HII region properties. Each pulsar-\HII region pair was validated against multi-wavelength maps, and about 30 spurious intersections have been excluded from an initial sample of $134$ candidates. Most of the spurious candidates correspond to \HII regions in HH14 with small near kinematic distances, leading to an inaccurately large angular diameter used for the impact parameter cutoff. We thus estimate a false positive rate of $\approx 20\%$ for the full set of $660$ intersections identified above. For some pulsars, our criteria yielded multiple \HII region intersections; these cases are noted in Table~\ref{tab:intersections1} but we only report properties for the \HII region closest to the pulsar LOS. 

\begin{figure}
    \centering
    \includegraphics[width=0.44\textwidth]{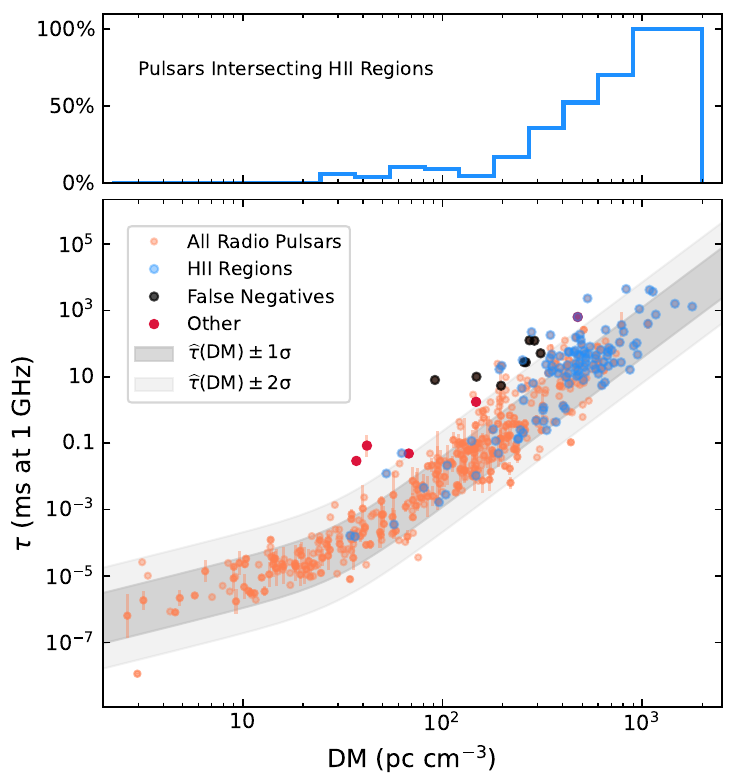}
    \caption{Pulse broadening time $(\tau)$ in ms at 1 GHz vs. DM for 500 pulsars (orange points) with scattering measurements compiled by this work and \cite{cordes2016,cordes2022}, where $\tau$ has been referenced to a frequency $\nu = 1$ GHz assuming $\tau \propto \nu^{-4}$. {In cases where multiple measurements were available, the mean and rms of those measurements are shown.} The grey and light grey bands show the $1\sigma$ ($68\%$) and $2\sigma$ ($95\%$) confidence intervals of the best-fit empirical $\tau$-DM relation from \cite{cordes2022}. Blue points indicate pulsars identified as intersecting \HII regions based on the criteria described in Section~\ref{sec:intersections}, and black points show false negatives, i.e., intersections that were missed due to the \HII region lacking a distance measurement or due to a misestimated pulsar distance. We only searched for false negatives among pulsars with $\tau$ more than $2\sigma$ larger than the best-fit $\tau$-DM relation. Red points indicate outliers with large scattering that intersect discrete structures other than \HII regions; these include pulsars within supernova remnants and pulsars intersecting the Gum Nebula and the Orion-Eridanus superbubble. The top panel shows the fraction of pulsars that intersect \HII regions, {in individually normalized DM bins}.}
    \label{fig:tau_vs_dm}
\end{figure}

To estimate a false negative rate for our intersection analysis, we focus on a subset of pulsars with scattering in excess of values typical for a given DM. When the observed scattering times of many pulsars are referenced to a single observing frequency $\nu$ (assuming, e.g., $\tau \propto \nu^{-4}$; \citealt{rickett1990}), they show a nonlinear, positive correlation with DM \citep{cordes91,bhat2004,krishnakumar15,cordes2016}. This empirical $\tau$-DM relation has the form \citep{cordes2022}: $\widehat{\tau}(1~\rm GHz) = (1.9\times10^{-7} \rm~ms)~DM^{1.5}(1 + 3.55\times10^{-5}DM^{3})$. Figure~\ref{fig:tau_vs_dm} shows the distribution of scattering times vs. DMs and the best-fit $\tau$-DM relation, which has a $68\%$ confidence interval $\sigma(\rm log_{10}~\tau) = 0.76$ \citep{cordes2022}. Twenty-six pulsars in our sample have $\tau$ more than $2\sigma$ greater than the $\tau$-DM relation (hereafter referred to as ``excess scattering''), with eight of these pulsars positively identified as intersecting \HII regions. Out of the remaining 18 pulsars with excess scattering, 8 {are spatially coincident in multi-wavelength maps} with \HII regions in the WISE and HH14 catalogs and appear to have been missed by the intersection analysis, due to a combination of the \HII regions lacking distance measurements and misestimated pulsar distances. Another 5 of the pulsars with excess scattering intersect known structures that are not \HII regions, including supernova remnants, the Gum Nebula, and the Orion-Eridanus superbubble, {based on assessment of their locations, angular extents, and estimated distances}. We are unable to identify known foreground structures for the 5 remaining pulsars with large scattering. The full list of pulsar scattering outliers and potentially relevant foreground structures is given in Appendix~\ref{app:outliers}; {the appendix also include cases where $\tau$ is over $2\sigma$ \textit{smaller} than the mean $\tau$-DM relation, which may be attributable to discrete structures close enough to the source that they contribute a DM enhancement but no significant scattering relative to the ISM \citep{cordes2016}}. 

We thus find that the majority of pulsars with excess scattering intersect a known discrete structure in the ISM, with more than half of those structures being \HII regions. Based on the \HII region intersections that were missed for these cases, we estimate the false negative rate to be between $\approx 8/500 = 2\%$ and $8/26 = 30\%$, where we regard the latter value as an upper limit because it is based on only those pulsars with excess scattering. 

Given the impact parameter cutoff used to identify \HII region intersections, it is likely that the false positive and negative rates estimated above correspond to \HII regions around specific star types. Many of the false positive intersections may correspond to B-type stars and later, which are typically smaller than the $20$ pc impact parameter cutoff assumed for the HH14 cross-match. Conversely, many of the missed intersections may correspond to early O-type stars, which can have sizes $>20$ pc \citep{rubin1968}. Despite their small sizes, \HII regions around B1/B2 stars (in addition to white dwarfs), may be sufficient in number to enhance the overall density of the warm ionized medium (WIM), particularly in the inner galaxy. We thus argue that the purity of the intersection sample is not critical to our subsequent analysis, because even the false positive intersections provide a diagnostic of the extent to which \HII regions can enhance the WIM.

\section{General Trends}\label{sec:trends}

\begin{figure*}
    \centering
    \includegraphics[width=0.9\textwidth]{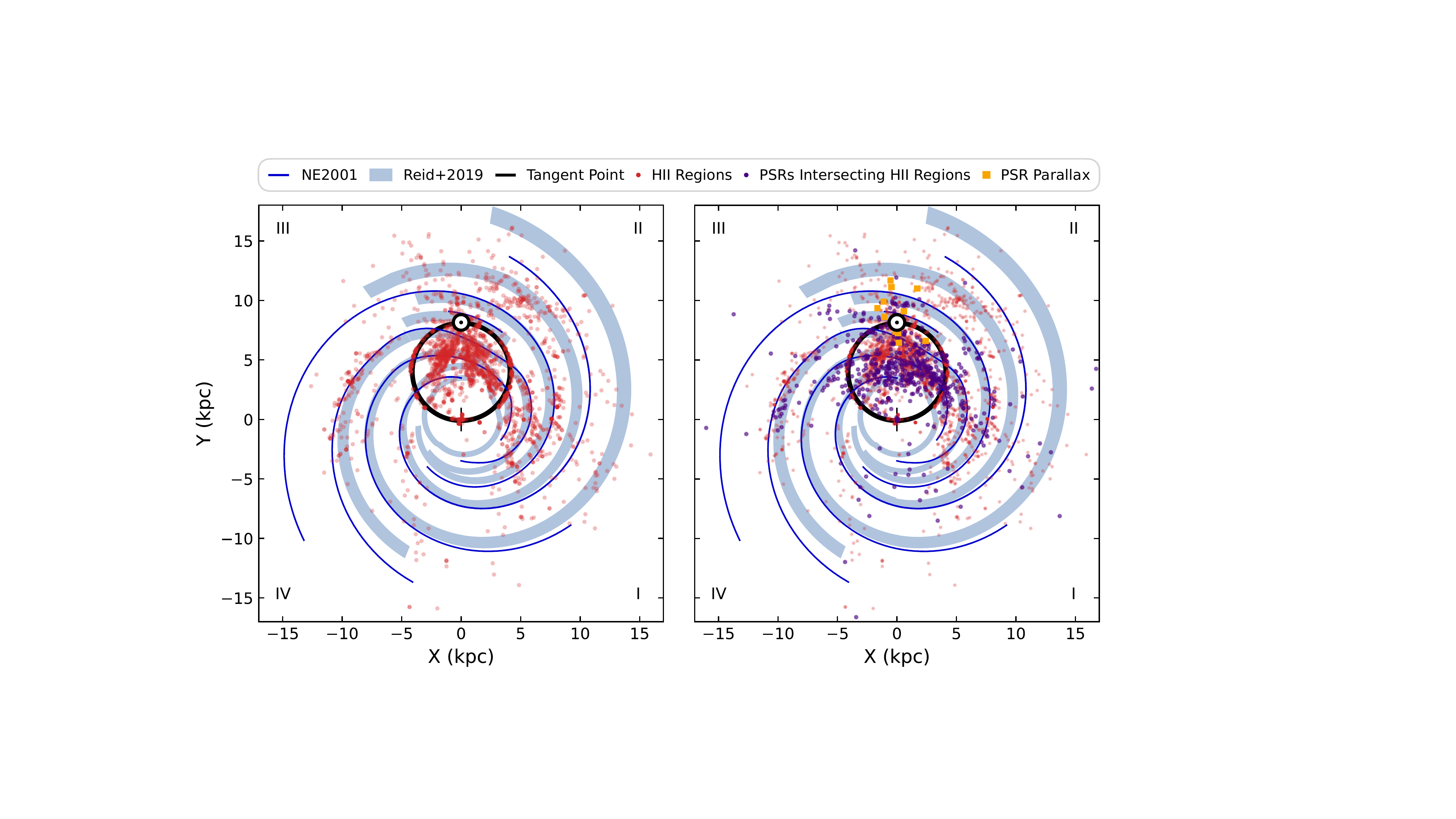}
    \caption{{Distribution of \HII regions and pulsar intersections projected onto the Galactic plane, with the Galactic Center at the origin.} \textit{Left:} {\HII regions with cataloged distances (red points)} are superposed on the spiral arm model used in NE2001 (blue curves) and the \cite{reid2019} best-fit model (light blue shaded curves). {Tangent point distances lie along the black circle.} Galactic quadrant numbers are noted in the corners. \textit{Right:} Pulsar sightlines identified as intersecting \HII regions are shown in purple. Most of the pulsar distances shown here are estimated using Galactic electron density models, {except for the orange squares, which indicate pulsars with parallax distances}. We thus expect that many of these pulsars, including those near the outskirts of the Galaxy, are closer to the observer than they appear in this figure.}
    \label{fig:spiral_arms}
\end{figure*}

We find that the majority of pulsars with large DM and $\tau$ intersect \HII regions. Figure~\ref{fig:dm_vs_lon} shows the fraction of pulsars intersecting \HII regions as functions of both DM and Galactic longitude, for all radio pulsars at Galactic latitudes $|b|<5^\circ$. Given that \HII regions are largely confined to the Galactic plane, we are interested in comparing the radio properties of pulsars intersecting \HII regions against the radio properties of other pulsars in the plane. For these low-latitude pulsars, we find that $>50\%$ of pulsars with $\rm DM > 580$ pc cm$^{-3}$ are identified as intersecting \HII regions. These intersections are roughly uniform across longitudes $|l|\lesssim50^\circ$, with notable clusters of intersections around $l \approx 150^\circ$, $l\approx 80^\circ$, $l \approx -100^\circ$, and $l \approx -175^\circ$ (likely caused by several prominent \HII regions, including S205, the \object[Cyg Complex]{Cygnus region}, S273, S264, and a series of \HII regions along the upper edge of the Gum Nebula). Figure~\ref{fig:dm_vs_lon} also shows the maximum DM predicted by NE2001 in the Galactic plane, demonstrating that most of the pulsars with $\rm DM > DM_{\infty}(NE2001)$ intersect \HII regions. The maximum DM predicted by YMW16 is generally larger than that of NE2001, but we similarly find that most of the pulsars with $\rm DM > DM_{\infty}(YMW16)$ lie behind \HII regions. 

For the pulsars with scattering measurements, $>50\%$ of pulsars with $\rm DM > 600$ pc cm$^{-3}$ and $\tau > 10$ ms at 1 GHz are identified as intersecting \HII regions. The fraction of these pulsars behind \HII regions is a steep function of DM that grows roughly as DM$^3$, which would be expected for an intersection probability that scales with the volume probed. Previous studies have linked small samples of both pulsars and extragalactic radio sources with large scattering to \HII region envelopes \citep{anantharamaiah1988} and regions of high H$\alpha$ emission \citep{spangler1990}, supporting interpretations of the $\tau$-DM relation that attribute the steepening of the relation at high DM to enhanced density fluctuations in the inner Galaxy \citep{cordes91,ocker2021halo}. {Interplanetary scintillation (IPS) of compact radio sources also appears to be systematically suppressed for LOSs through regions of enhanced H$\alpha$ emission \citep{morgan2022}, further suggesting that \HII regions generate enough scattering to quench IPS.} {Complementary assessments of Faraday rotation measures (RMs) have shown that RM variations in the Local and Carina spiral arms are consistent with a turbulence injection scale $\approx 2$ pc \citep{haverkorn2004}, suggesting that \HII regions play a role in the turbulence cascade that gives rise to pulsar scattering.} Our results support this general picture, but \HII regions do not appear to be solely responsible for the steepening of the $\tau$-DM relation, as evidenced by the many pulsars between $100 \lesssim \rm DM \lesssim 600$ pc cm$^{-3}$ that are not in the \HII region sample (see Figure~\ref{fig:tau_vs_dm}). However, some of these moderate-DM pulsars are likely missed intersections, in part because there are many more \HII regions with inferred distances in the inner Galaxy. 

Figure~\ref{fig:spiral_arms} shows the locations of \HII regions with distances in the WISE and HH14 catalogs, along with the pulsar intersections identified in Section~\ref{sec:candidates}, projected onto the Galactic plane. Several selection effects are evident in this figure, and demonstrate the sources of observational bias inherent in our results. First, there are many more \HII regions with inferred distances in the inner Galaxy of quadrant I. This asymmetry in the distribution of \HII region distances is partially due to radio telescope coverage at higher sensitivity, leading to a more complete distribution of radial velocities and distances in quadrant I \citep{anderson2012}. A similar asymmetry manifests in the distribution of pulsars because of the northern PALFA pulsar survey \citep{cordes2006_palfa,parent2022_palfa}. Uncertainties in the \HII region distances are complex, depending on the inference method used and the \HII region location. Kinematic distances in the inner Galaxy are highly uncertain when they correspond to radial velocities near the maximum velocity of the inner Galaxy. These tangent point distances are highlighted in Figure~\ref{fig:spiral_arms} and appear as a circular artifact when projected onto the Galactic plane. The most precise distances for \HII regions are based on maser parallaxes and are almost exclusively in quadrants I and II \citep{reid2019}. Taken together, these various selection effects suggest that false positive intersections in our analysis are predominantly for LOSs in the inner Galaxy, where kinematic distance uncertainties are large, and that more intersections are missed (false negatives) in quadrants III and IV, where the sample of \HII region distances is less complete. Figure~\ref{fig:spiral_arms} also shows spiral arm models from NE2001 and \cite{reid2019}; these models are discussed further in Section~\ref{sec:implications}.

\section{Comparison to Radio Recombination Lines}\label{sec:gdigs_fast}

\begin{figure*}
    \centering
    \includegraphics[width=\textwidth]{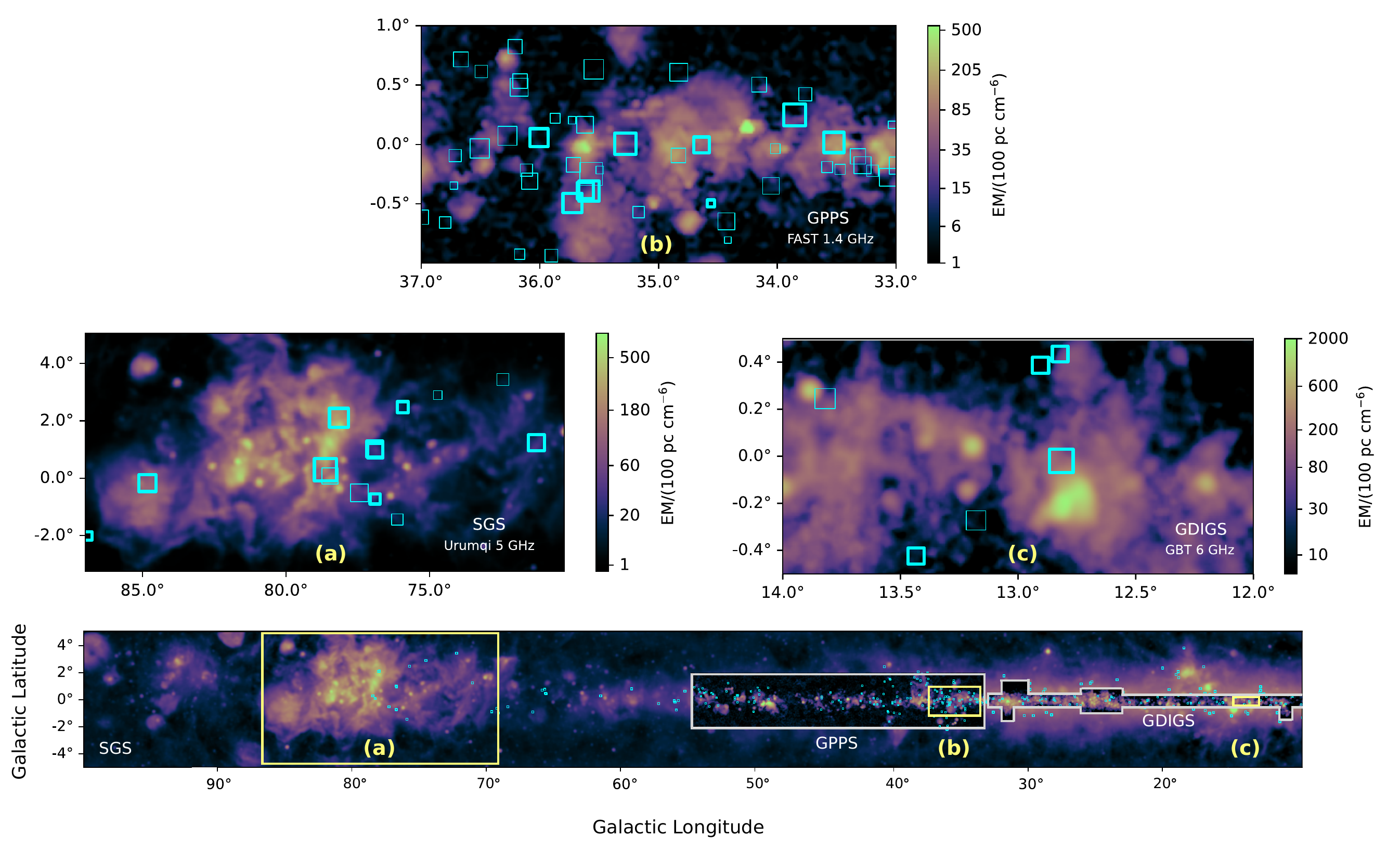}
    \caption{{Maps of emission measure (EM) in units of 100 pc cm$^{-6}$ for three sky areas containing \HII regions along pulsar LOSs (cyan squares). The bottom panel demonstrates the regions selected from each survey for comparing \HII region EMs to pulsar DMs (yellow squares). GDIGS and SGS respectively extend to smaller and larger longitudes than shown here. Panel (a) shows EM estimated from total intensity measured by SGS \citep{sun2007} at Urumqi, assuming an RRL-to-continuum ratio of $13\%$ and a 25 km s$^{-1}$ line width (\S\ref{sec:em_maps}). Panels (b) and (c) show EM estimated from Hn$\alpha$ intensity maps measured by the GPPS \citep{hou2022_fast} and GDIGS \citep{anderson2021_gdigs} surveys. The sizes of the cyan squares indicate the relative magnitudes of the pulsars' DMs (larger squares mean larger DMs). Thicker squares are pulsars with scattering measurements.}}
    \label{fig:gdigs_gpps_sino}
\end{figure*}

Having found that the majority of pulsars with large DM and $\tau$ lie behind \HII regions, we now must ask: How much of their DM and scattering is contributed by the \HII regions vs. the rest of the ISM along the LOS? Recombination lines provide an independent observational constraint on the electron density content of these regions via estimation of their emission measures (${\rm EM} = \int n_e^2 dl$). Many of the \HII regions in our sample do not have optical recombination line observations at the spatial resolution needed for even a broad comparison to pulsars in our sample. We hence turn to radio surveys of \HII regions, which have the added benefit of not suffering from dust extinction. We select three areas on the sky that contain a few dozen pulsars behind \HII regions, and that have been covered by radio surveys with sufficient sensitivity and resolution to estimate EMs near the pulsar LOSs for comparison to DM and scattering. {These three sky areas are shown in Figure~\ref{fig:gdigs_gpps_sino}, using data from subsets of three radio surveys}: the Green Bank Diffuse Ionized Gas Survey (GDIGS; \citealt{anderson2021_gdigs}), FAST GPPS \citep{hou2022_fast}, and the Sino-German 6 cm Survey (SGS) \citep{sun2007,sun2011,xiao2011,gao2019}. Table~\ref{tab:surveys} gives an overview of the three surveys. {While the total sky coverage of the surveys includes nearly half of the Galactic plane, we focus on the three areas shown in Figure~\ref{fig:gdigs_gpps_sino} because they each contain multiple pulsars where the \HII regions' radio emission is bright enough to accurately infer EMs.} 

\begin{deluxetable*}{c c c C C c c}\label{tab:surveys}
\tabletypesize{\footnotesize}
\tablecaption{Radio Surveys of \HII Regions}
\tablehead{\colhead{Survey} & \colhead{Telescope} & \colhead{Observing Frequency} & \colhead{Sky Coverage} & \colhead{Angular Resolution} & \colhead{Median $W_{\rm RRL}$ Uncertainty} & \colhead{EM Sensitivity}}
\startdata
GDIGS & GBT & 4--8 GHz & 32.3^\circ > l > -5^\circ, |b|<0.5^\circ & 2.65^{\prime} & 0.26 K km/s & 1100 pc cm$^{-6}$ \\
GPPS & FAST & 1--1.5 GHz & 55^\circ > l > 33^\circ, |b|<2^\circ & 3^\prime & 0.16 K km/s & 200 pc cm$^{-6}$ \\
SGS & Urumqi & 4.5--5.1 GHz & 10^\circ < l < 230^\circ, |b|<5^\circ & 9.^\prime5 & -- & -- \\
\enddata
\tablecomments{Overview of the three radio surveys used in this work: the Green Bank Diffuse Ionized Gas Survey (GDIGS; \citealt{anderson2021_gdigs}), the FAST Galactic Plane Pulsar Survey (GPPS; \citealt{hou2022_fast}), and the Sino-German 6 cm Survey (SGS; \citealt{gao2019}. The sky coverages shown correspond to the total survey footprints, of which we only use a portion in our analysis. {The rightmost two columns give the median uncertainty in RRL line intensity assuming a 25 km/s line width, and an equivalent EM sensitivity for $T_e = 8000$ K and a $3\sigma$ detection threshold.}
Both GDIGS and GPPS resolve RRLs (Hn$\alpha$, Hn$\beta$, Hn$\gamma$) and provide maps of Hn$\alpha$ integrated intensity. SGS does not resolve RRLs but has a total brightness temperature sensitivity of $\approx 1$ mK, comparable to the sensitivity of GDIGS when integrating over all Hn$\alpha$ lines.}
\end{deluxetable*}

\subsection{Emission Measure Maps}\label{sec:em_maps}

GDIGS and FAST GPPS both have the spectral resolution needed to distinguish the Hn$\alpha$ recombination line transition (in addition to Hn$\beta$ and Hn$\gamma$) and they provide velocity-integrated Hn$\alpha$ intensity maps in their respective frequency ranges. In local thermodynamic equilibrium, the velocity-integrated line intensity $W_{\rm RRL}$ is related to EM as \citep{rrl_textbook,anderson2021_gdigs}:
\begin{equation}\label{eq:rrl_to_em}
    \frac{\rm EM}{\rm pc\ cm^{-6}} \approx 1150 \bigg(\frac{\nu}{\rm GHz}\bigg) \bigg(\frac{T_e}{8000\ \rm K}\bigg)^{3/2} \bigg(\frac{W_{\rm RRL}}{\rm K\ km\ s^{-1}}\bigg),
\end{equation}
where $T_e$ is the electron temperature and $\nu$ is the frequency of the line transition. The pre-factor in Equation~\ref{eq:rrl_to_em} applies only to Hn$\alpha$ lines (i.e., an energy level transition $\Delta n = 1$; see Appendix A of \citealt{anderson2021_gdigs}). \HII regions typically have $T_e \approx 8000$ K, although inferred values can range from $\approx 5,000 - 20,000$ K (equivalent to up to a factor of 4 change in estimated EM) {and tend to increase with Galactocentric radius \citep{quireza_2006_1,khan2024}. We use the best-fit form of $T_e$ as a function of Galactocentric radius $R_{\rm gal}$ from \cite{quireza_2006_1}, 
\begin{equation}\label{eq:Te_gradient}
    T_e = (5780\pm350) + (287\pm46)R_{\rm gal}
\end{equation}
and we estimate $T_e$ for \HII regions in each of the sky areas we examine. To estimate the resulting EM and its uncertainty, we sample from normal probability density functions (PDFs) for both $T_e$ and the RRL intensity $W_{\rm RRL}$, based on the $68\%$ confidence intervals of the $T_e$ gradient and the rms spectral noise of the RRL intensity maps, respectively.}

SGS does not provide velocity-integrated Hn$\alpha$ maps, and so we convert from total intensity to EM by assuming an Hn$\alpha$ line width of 25 km s$^{-1}$ and a line-to-continuum ratio ($R_{\rm LC}$; the ratio of RRL intensity to the total radio intensity integrated over the total frequency bandwidth). 
SGS is one of the most sensitive radio surveys with contiguous coverage of Galactic longitudes $90^\circ > l > 70^\circ$, which is a region of high interest because it contains a large ($\approx 90$ deg$^2$) complex of \HII regions and the pulsars identified both by \cite{han2021_fast} and this work as having DMs significantly greater than Galactic model predictions. This sky area (which includes the famous Cygnus region) has only been partially covered by previous RRL surveys, including GDIGS and \cite{barcia1985}, \cite{heiles1996}, and \cite{azcarate1997}. We leverage GDIGS Hn$\alpha$ maps covering a portion of Cygnus to estimate $R_{\rm LC}$, finding it to be as low as $4\%$ in the faintest regions and as high as $30\%$ in the brightest, with the majority of locations in the range of $10-15\%$. These ratios are entirely consistent with empirical ratios for dozens of other \HII regions \citep{quireza2006_2}. We therefore adopt a mean value $R_{\rm LC} = 13\%$, and for subsequent analysis we adopt a log-normal PDF for $R_{\rm LC}$ spanning the range of values given above.

Figure~\ref{fig:gdigs_gpps_sino} shows maps of the EMs inferred from {subsets of} these three surveys using Equation~\ref{eq:rrl_to_em}, based on the measured line intensities and {median value of $T_e$ in each sky area}. The smallest EMs inferred are comparable to the sensitivities listed in Table~\ref{tab:surveys}, but are dramatically larger ($>10^4$ pc cm$^{-6}$) near the centers of the \HII regions. {Table~\ref{tab:surveys} gives the median $W_{\rm RRL}$ uncertainty for each survey, which combined with uncertainties in $T_e$ (due to both scatter in the $T_e$ gradient and uncertainty in the \HII region distances) yields typical EM uncertainties $\approx 30 - 60\%$.} The spectral noise in all of these survey maps varies spatially by up to $\approx 200\%$, with the highest noise levels corresponding to regions of highest radio intensity. For SGS, we do not have velocity-resolved spectral rms measurements. However, the total brightness temperature sensitivity reported for SGS by \cite{xiao2011}, 1 mK, is very similar to the brightness temperature sensitivity of GDIGS after integrating over all Hn$\alpha$ lines \citep{anderson2021_gdigs}. {We consequently assume that the equivalent $W_{\rm RRL}$ uncertainty is similar, and add it in quadrature to the uncertainty associated with $R_{\rm LC}$.}

\subsection{Relating Emission Measure to Dispersion Measure}\label{sec:EM_to_DM}

\begin{figure*}
    \centering
    \includegraphics[width=0.7\textwidth]{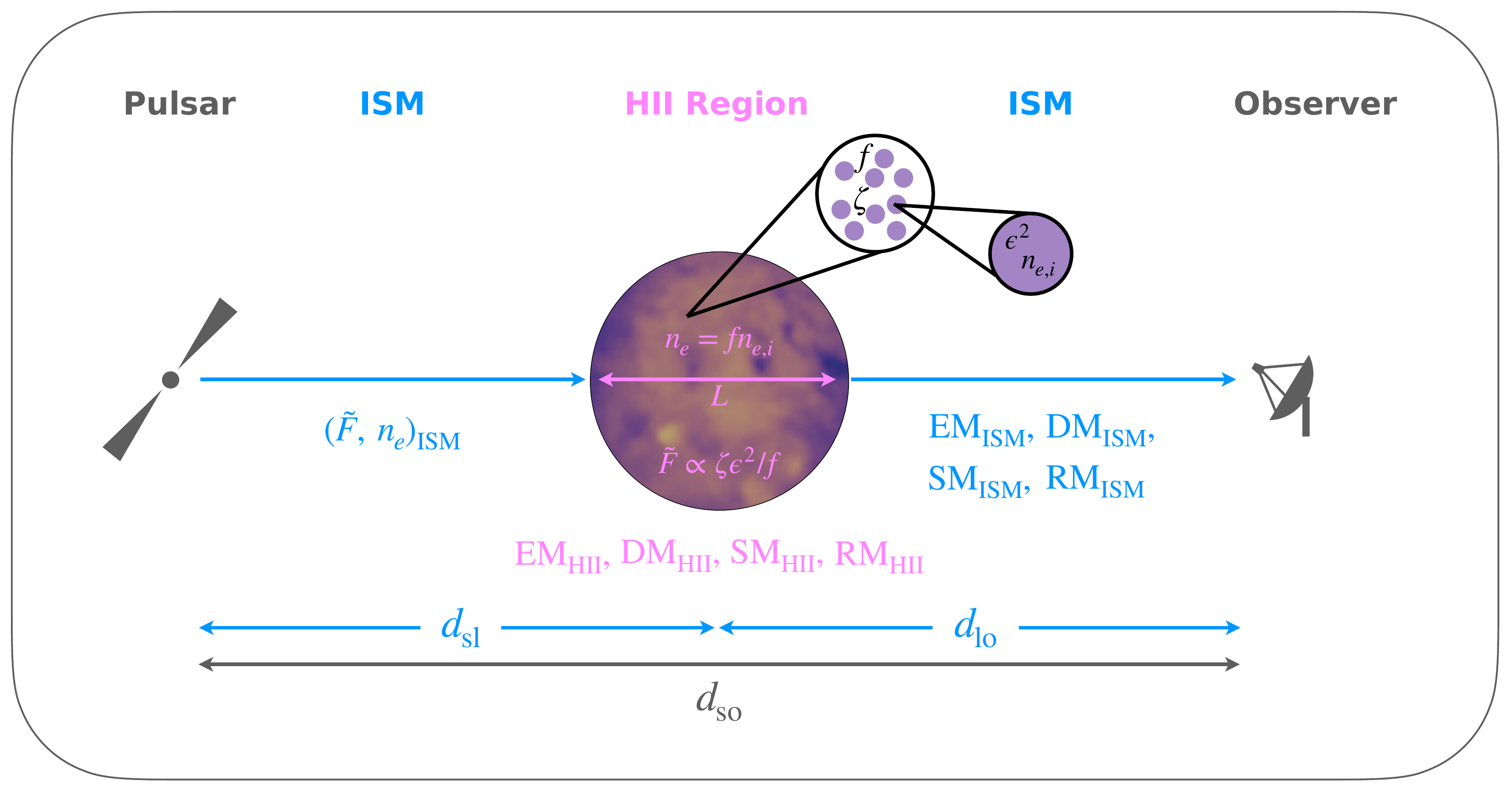}
    \caption{{Schematic of the ionized cloudlet model used to relate pulsar and emission line observables (\S\ref{sec:EM_to_DM}-\ref{sec:scattering}). \HII regions are modeled with electron densities $n_e$ and density fluctuation parameters $\tilde{F}$. The fluctuation parameter depends on the interior properties of the \HII region, described by the volume filling factor of cloudlets $f$, the density variance within cloudlets $\epsilon^2$, and density variations between cloudlets $\zeta$. The ISM is described by separate $n_e$ and $\tilde{F}$. The relevant observables are emission measure (EM), dispersion measure (DM), scattering measure (SM, which is related to the scattering delay $\tau$), and rotation measure (RM). In this work we only consider relations between EM, DM, and $\tau$.}}
    \label{fig:cloudlet_schematic}
\end{figure*}

We relate EM to DM using an ionized cloudlet model, in which cloudlets have a local volume-averaged mean density~$n_e$ and internal density~$n_{e,i}$, such that $n_e = f n_{e,i}$ for a volume filling factor~$f$. The electron density fluctuation variance within a cloudlet is $\epsilon^2 = (\langle (\delta n_{e,i})^2 \rangle / n_{e,i}^2$, and variations between cloudlets $\zeta = \langle n_{e,i}^2 \rangle / \langle n_{e,i} \rangle^2$ \citep{cordes91,1993ApJ...411..674T,ne20011}. {Figure~\ref{fig:cloudlet_schematic} shows a schematic of the model.} The resulting relation between EM and DM is
\begin{equation}\label{eq:dm_from_em1}
    {\rm EM} = \frac{\zeta(1+\epsilon^2)}{fL}{\rm DM^2},
\end{equation}
where $L$ is the path length through the relevant volume of cloudlets---in our case, the path length through an \HII region (see also Appendix~B of \citealt{cordes2016}). Here $\zeta$, $\epsilon$, and $f$ are dimensionless quantities, and $L$ and DM have their usual units of pc and pc cm$^{-3}$. In practice, $\zeta \geq 1$, $\epsilon^2 \leq 1$, and $f\leq 1$, and we can place an upper limit on DM for a given EM, $f$, and $L$ by assuming $\zeta \rightarrow 1$ and $\epsilon^2 \rightarrow 0$ {(corresponding to the limit of no turbulent fluctuations within cloudlets)}:
\begin{equation}\label{eq:dm_from_em2}
    {\rm DM} \leq 100\ {\rm pc\ cm^{-3}}\times (f_{0.1} L_{10} {\rm EM}_{4})^{1/2}
\end{equation}
where $f_{0.1} = f/0.1$, $L_{10} = L/(10\ {\rm pc})$, and $\rm EM_4 = EM/(10^4$ pc cm$^{-6})$. The equivalent volume-averaged mean electron density would be $n_e = {\rm DM}/L$, or
\begin{equation}
    n_e \leq 10\ {\rm cm^{-3}}\ (f_{0.1} {\rm EM_4}/L_{10})^{1/2}.
\end{equation}

\begin{figure}
    \centering
    \includegraphics[width=0.47\textwidth]{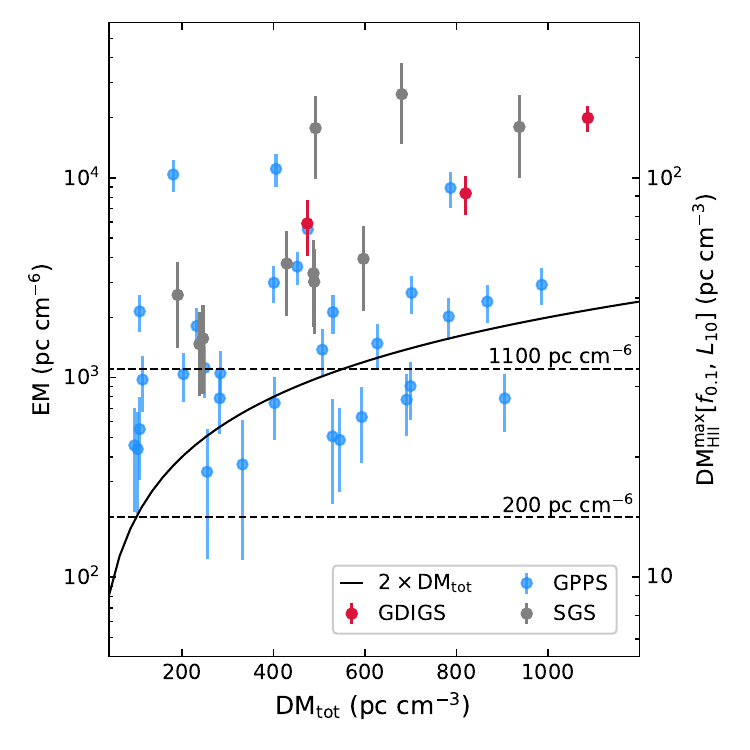}
    \caption{EM vs. total observed DM for pulsars in the three sky regions shown in Figure~\ref{fig:gdigs_gpps_sino}. Pulsars covered by a portion of the GDIGS survey are shown in red, those covered by a portion of GPPS in blue, and those covered by a portion of SGS in grey. EMs were calculated using the procedure given in Sections~\ref{sec:em_maps}-\ref{sec:EM_to_DM}. {Error bars account for the uncertainties of both $W_{\rm RRL}$ (Table~\ref{tab:surveys}) and $T_e$ (Section~\ref{sec:em_maps}).} For the SGS data we assume the same median $W_{\rm RRL}$ uncertainty as for GDIGS, and we sample from a log-normal PDF for $R_{\rm LC}$ to include its uncertainties in the error bars shown. {Pulsar LOSs with EMs smaller than the survey's EM sensitivity and fractional EM errors $>75\%$ have been excluded.} The right-hand axis shows the equivalent upper limit on DM assuming a volume filling factor $f = 0.1$ and a path length through the \HII region $L = 10$ pc (see Section~\ref{sec:EM_to_DM}). {The black curve indicates $\rm EM = 2\ \rm DM_{tot}$ (equivalently, $\rm DM_{HII}^{max} = [2\ DM_{tot}]^{1/2}$) to guide the eye.}}
    \label{fig:em_to_dm}
\end{figure}

Figure~\ref{fig:em_to_dm} shows the distribution of EM vs.\ DM for the pulsars covered by {each cutout of} \hbox{GDIGS}, \hbox{GPPS}, and SGS. {The EM is estimated for each pulsar LOS using the nearest map pixel} and evaluating $T_e$ for the Galactocentric radius of the intersected \HII region (Equation~\ref{eq:Te_gradient}). There is only weak correlation (Pearson's coefficient $r = 0.4$) between EM and total observed DM, which is unsurprising given that the total DMs include contributions from ionized gas in the diffuse ISM, and given that variations in $f$, $L$, $\zeta$, and $\epsilon$ between different regions are guaranteed.  Estimating the DM contribution from the diffuse ISM alone is fraught in the absence of distance measurements for these pulsars, which would enable rough tomography of these sky regions. We instead focus on estimating the range of possible DMs that could be contributed by the \HII regions ($\rm DM_{HII}$) based on their EMs. Figure~\ref{fig:em_to_dm} shows the maximum DM estimated from EM based on Equation~\ref{eq:dm_from_em2} for the pulsars in the GDIGS, GPPS, and SGS fields, assuming $f = 0.1$ and $L = 10$ pc, which are considered typical for $10^4$\,K ionized gas and \HII regions \citep{draine2011}. The resulting upper limits on $\rm DM_{HII}$ range from $\approx 15$ to~200\,pc\,cm$^{-3}$, with median values $\approx 10\%$ of the total observed \hbox{DM}.

These fiducial EM-inferred DMs may underestimate the DM contributions of the \HII regions, particularly for those pulsars which have total DMs greater than Galactic models' maximum predictions. For example, PSR J2030$+$3944g has a DM excess of 540 pc cm$^{-3}$ greater than $\rm DM_{\infty}$ predicted by NE2001, but the fiducial value ${\rm DM_{HII}^{max}}(f = 0.1, L = 10) \approx 180$ pc cm$^{-3}$ suggests that $f$ and/or $L$ must be larger for the \HII region to account for the pulsar's excess DM.

We examine this further by estimating PDFs for $\rm DM_{HII}$ through a likelihood analysis of each pulsar with a total DM in excess of $\rm DM_\infty(NE2001)$. The per-pulsar PDF is
\begin{multline}
    p({\rm DM_{HII} | EM}, F_\ell)
     \sim \int d {\rm EM}\ d F_\ell \ p({\rm EM}) p(F_\ell)\\ \times \delta [{\rm DM_{HII}} - \mathcal{G}({\rm EM}, F_\ell)],
\end{multline}
where $F_\ell = fL/\zeta(1+\epsilon^2)$ and $\mathcal{G}({\rm EM}, F_\ell)$ is the function relating DM and EM (Equation~\ref{eq:dm_from_em1}). The PDF for EM is evaluated using priors on $T_e$, $W_{\rm RRL}$, and $R_{\rm LC}$ described in Section~\ref{sec:em_maps}, and we adopt a flat prior for $F_\ell$ restricted to the range $[0.1,50]$ pc, which includes cases spanning $f/\zeta(1+\epsilon^2) \ll 1$ to $f/\zeta(1+\epsilon^2) \sim 1$ and $L \sim 5$ to 50 pc. We then compare $p(\rm DM_{HII})$ to $\rm DM_{excess} = DM_{tot} - DM_{\infty}(NE2001)$, the pulsar's excess from the maximum Galactic DM prediction for that LOS, and use a numerical joint likelihood analysis of $p(\rm DM_{HII})$ and $\rm DM_{excess}$ to yield a posterior PDF for $F_\ell$. In short, we search for the range of $F_\ell$ that allows the \HII region's EMs to account for each pulsar's entire excess DM.

\begin{figure}
    \centering
    \includegraphics[width=0.47\textwidth]{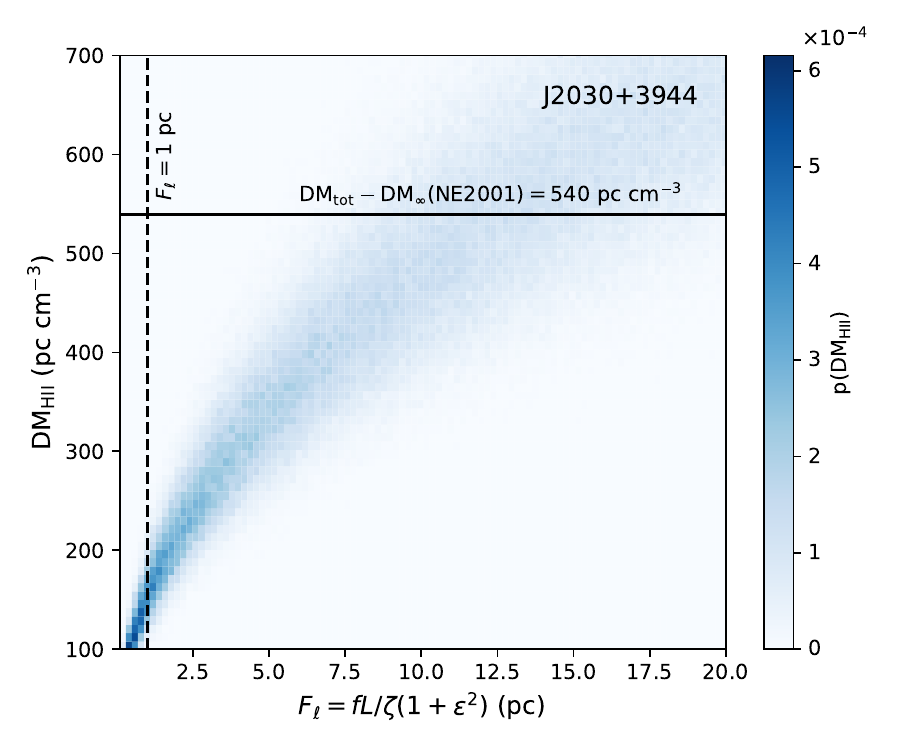}
    \caption{Probability density for $\rm DM_{HII}$ vs.\ $F_\ell$ for PSR~J2030$+$3944g, calculated via the likelihood analysis described in Section~\ref{sec:EM_to_DM}, which assumes a flat prior on $F_\ell$. The black horizontal line indicates the pulsar's excess DM from NE2001's maximum DM for that \hbox{LOS}, and the vertical dashed line indicates the fiducial value $F_\ell = 1$, which would yield the upper limit $\rm DM_{HII}^{max}$ for $f = 0.1$, $L = 10$\,pc, and $\zeta = 1+\epsilon^2 = 1$, as assumed in Figure~\ref{fig:em_to_dm}. In order for the \HII region's EM to be consistent with the entire excess \hbox{DM}, the PDF for $\rm DM_{HII}$ requires $F_\ell = 13^{+5}_{-4}$\,pc. {This constraint on $F_\ell$ is a lower limit because $\rm DM_{\infty}(NE2001)$ can be considered an upper bound on the ISM contribution to the pulsar's DM.}}
    \label{fig:dm_h2_grid}
\end{figure}

Figure~\ref{fig:dm_h2_grid} shows an example of $p(\rm DM_{HII})$ for PSR J2030$+$3944g, which has the largest excess DM in our sample, compared to the value of $\rm DM_{HII}$ that would be inferred assuming $F_\ell = 1$ pc, which is equivalent to the upper limit $\rm DM_{HII}^{max}$ for $f = 0.1$, $L = 10$ pc, and $\zeta = 1+\epsilon^2 = 1$. In order for the \HII region's EM to account for the entire excess DM, $F_\ell$ must be larger, and we find $F_\ell = 13^{+5}_{-4}$ pc (corresponding to the $50\%^{+35\%}_{-35\%}$ confidence intervals of the posterior PDF). We perform this analysis for all of the pulsars covered by the EM maps in Figure~\ref{fig:gdigs_gpps_sino} with DMs in excess of the predicted maximum DM, which yields constraints on $F_\ell$ for six LOSs shown in Table~\ref{tab:fL}. Half of these pulsars have $F_\ell \sim 1$ to $5$ pc, broadly consistent with standard assumptions of $f$ and $L$ for the case $\zeta = 1+\epsilon^2 = 1$. In one case, J2030$+$3929g, $F_\ell < 1$ and the EM is similar in value to the pulsars with $\rm DM_{excess} > 300$ pc cm$^{-3}$, which suggests that the \HII region may account for even more of the pulsar's DM. In principle, these constraints on $F_\ell$ {could be considered} lower limits because $\rm DM_{\infty}(NE2001)$ provides a {model-dependent} upper bound on the ISM contribution to DM along the pulsar LOS, and hence $\rm DM_{excess}$ (and $F_\ell$) could be even larger.

\begin{deluxetable}{c C C C}\label{tab:fL}
\tabletypesize{\footnotesize}
\tablecaption{Constraints on $F_\ell$ for Pulsars with $\rm DM_{tot} > DM_{\infty}(NE2001)$}
\tablehead{\colhead{PSR} & \colhead{$\rm DM_{excess}$ (pc cm$^{-3}$)} & \colhead{EM ($10^3$ pc cm$^{-6}$)} & \colhead{$F_\ell$ (pc)} }
\startdata
J2005$+$3411g & 73 & 3.6^{+1.4}_{-1.0} & 1.5^{+0.5}_{-0.4} \\
J2030$+$3929g & 91 & 21.1^{+8.6}_{-5.9} & 0.4^{+0.1}_{-0.1} \\
J2022$+$3845g & 94 & 3.9^{+1.5}_{-1.0} & 2.3^{+0.9}_{-0.6} \\
J2030$+$3818g & 194 & 4.5^{+1.6}_{-1.2} & 8.5^{+3.1}_{-2.3} \\
J2021$+$4024g & 331 & 32.2^{+11.1}_{-8.6} & 3.4^{+1.2}_{-0.9}\\
J2030$+$3944g & 540 & 21.8^{+8.1}_{-5.9} & 13^{+5}_{-4} \\
\enddata  
\tablecomments{Values shown for $F_\ell$ {correspond to lower limits} based on the requirement that the \HII region EM be consistent with the entire DM excess from $\rm DM_{\infty}(NE2001)$. Error bars correspond to the $15\%$ and $85\%$ confidence intervals, and include uncertainties in $W_{\rm RRL}$, $T_e$, and $R_{\rm LC}$.}
\end{deluxetable}

Our results suggest that for naive assumptions about $f$ and $L$, the EMs inferred from RRL surveys can easily produce DM contributions $\sim 10 - 10^2$ pc cm$^{-3}$ from \HII regions. Allowing for an agnostic range of $f$, $L$, $\zeta$, and $\epsilon$ can increase the DM contributions of these \HII regions to several hundreds of pc cm$^{-3}$, which explains those pulsars with DMs far greater than the maximum model predictions for their LOSs. 

{All of the pulsars shown in Table~\ref{tab:fL} are in the vicinity of the Cygnus OB1 association. Previous assessments of extragalactic radio sources have found that Cygnus is a source of substantial scattering \citep{spangler1998} and variations in Faraday RMs as large as $\sim 100$s of rad m$^{-2}$ \citep{lazio1990,whiting2009}. While NE2001 includes about 10 clumps that reproduce the large scattering inferred from these prior studies, our assessment demonstrates that the DM contribution of Cygnus is up to $\sim 100\times$ larger than the model's clump parameters ($\approx 10 - 40$ pc cm$^{-3}$; \citealt{ne20011}), and a revised treatment of Cygnus will be needed in a next-generation Galactic electron density model.}

\section{Scattering and Density Fluctuations}\label{sec:scattering}

We use radio scattering measurements to constrain the density fluctuation properties of \HII regions through an extension of the ionized cloudlet model introduced in Section~\ref{sec:gdigs_fast}. We initially assume that the \HII regions are the dominant source of the pulsars' observed scattering (as supported by numerous studies that find radio wave scattering consistent with arising from \HII regions; e.g., \citealt{spangler1998,rickett2009,mall2022,ocker2024_scintarcs}). Potential complications from this assumption are discussed at the end of this section.

\subsection{Extension of the Ionized Cloudlet Model}

The ionized cloudlet model introduced in Section~\ref{sec:EM_to_DM} can be applied to the phenomenon of radio wave scattering by assuming a power spectrum of density fluctuations that extends down to spatial scales smaller than the Fresnel scale ($r_F \sim \sqrt{\lambda D} \sim 0.01$ au for typical distances $D$ and observing wavelengths $\lambda$). {Our implementation of the cloudlet model provides an analytic parameterization of scattering that is simple to constrain and comparable to the prescription adopted by NE2001. While we assume that pulse broadening arises from a Kolmogorov spectrum of turbulent density fluctuations, which is likely applicable to many but not all LOSs \citep[e.g.][]{cordes1986,oswald2021_meerkat}, the cloudlet model could be generalized to include scattering through non-turbulent structures.}
%While our implementation of the model is not necessarily an accurate description of the physical origin of scattering for all LOSs (see e.g., mixed evidence for consistency between Kolmogorov turbulence and pulsar scattering in \citealt{oswald2021_meerkat}, \citealt{ocker2024_scintarcs}, and refs. therein), it does provide an analytic parameterization of scattering that is simple to constrain and comparable to the prescription adopted by NE2001. The ionized cloudlet model can be generalized to include scattering through non-turbulent structures.}

We are primarily interested in the interpretation of the pulse broadening measurements shown in Figure~\ref{fig:tau_vs_dm}, which characterize the frequency-dependent broadening of radio pulses in time due to multipath propagation. For a statistically homogeneous medium and a Kolmogorov spectrum of density fluctuations of the form $P_{\delta n_e} = C_{\rm n}^2 q^{-11/3} {\rm exp}(-(q l_{\rm i}/2\pi)^2)$, the pulse broadening time in Euclidean space is \citep{cordes2016,ocker2021halo,cordes2022}:
\begin{equation}\label{eq:tau-dm-eq}
    \tau({\rm DM},\nu) \approx 48.03\ {\rm ns}\ A_\tau \nu_{\rm GHz}^{-4} \tilde{F} G_{\rm scatt} {\rm DM}^2.
\end{equation}
The prefactor given in Equation~\ref{eq:tau-dm-eq} for $\tau({\rm DM},\nu)$ is evaluated for $\nu$ in GHz and DM in pc cm$^{-3}$. The dimensionless factor $A_\tau \leq 1$ converts the mean pulse broadening time to the $1/e$ time that is typically estimated from pulse shapes. The exact value of $A_\tau$ depends on unknown properties of the scattering medium, and we proceed by assuming $A_\tau = 1$, which corresponds to the exponential scattering tail of a Gaussian scattered image. 

The density fluctuation spectrum introduced above depends on $C_{\rm n}^2$, the spectral amplitude, and $q = 2\pi/l$, the wavenumber for fluctuations at a given spatial scale $l$, where the spectrum extends from $2\pi/l_{\rm o} \lesssim q \lesssim 2\pi/l_{\rm i}$ for outer and inner scales $l_{\rm o}, l_{\rm i}$. For an \HII region, the outer scale is likely similar to the scale of the ionizing star's interaction with the ISM (comparable to the Str{\"o}mgren sphere radius, $\sim$10s of pc for O-type stars). The inner scale will be orders of magnitude smaller. In one case, \cite{rickett2009} measured the pulse broadening shape of the pulsar J1644$-$4559, which lies behind \HII region G339.1$-$0.4 (see Appendix~\ref{app:h2list}), at high enough precision to distinguish between different turbulence models, and found $l_{\rm i} \approx 70-100$ km. The values of $l_{\rm o}$ and $l_{\rm i}$ likely vary between \HII regions, so we combine these and the other unknown quantities describing the density fluctuation statistics of the medium into a composite density fluctuation parameter given by 
\begin{multline}\label{eq:ftilde}
    \tilde{F} \approx 0.08\ {\rm (pc^2 km)^{-1/3}}\ \frac{\zeta\epsilon^2}{f} \\ \times \bigg(\frac{l_{\rm o}}{20\ \rm pc}\bigg)^{-2/3} \bigg(\frac{l_{\rm i}}{100\ \rm km}\bigg)^{-1/3}.
\end{multline}

The geometric leverage factor $G_{\rm scatt}$ is derived from the standard Euclidean weighting of $C_{\rm n}^2$ and is given by (see \citealt{cordes2022}):
\begin{equation}\label{eq:gscatt}
    G_{\rm scatt}(x,y) = \frac{1 - (2x/3) + (2y/x)(1 - y - x)}{1 - (2x/3)}
\end{equation}
where $x = L/d_{\rm so}$ and $y = d_{\rm sl}/d_{\rm so}$ for a scattering layer of thickness $L$ at a distance $d_{\rm sl}$ from the source and a source-to-observer distance $d_{\rm so}$. 

The empirical $\tau$-DM relation observed for Galactic pulsars (Figure~\ref{fig:tau_vs_dm}) can be interpreted through this model as the result of variations in both $\tilde{F}$ and $G_{\rm scatt}$: LOSs towards the inner Galaxy have systematically higher $\tilde{F}$, leading to a different slope of the $\tau$-DM relation at high DMs, and LOSs at a given DM have different $G_{\rm scatt}$, leading to broad spread in $\tau$ for a fixed DM \citep{cordes2022}. If we simplistically assume that all pulsars are embedded in their scattering media (i.e., scattering is dominated by the extended ISM), then $G_{\rm scatt} \approx 1/3$ and the resulting distribution of $\tilde{F}$ varies by over 5 orders of magnitude between the inner and outer Galaxy, and as a function of Galactic latitude \citep{ocker2021halo}. Conversely, if scattering is dominated by discrete structures, then $G_{\rm scatt} > 1$ and we must explicitly account for how those structures are distributed along pulsar LOSs.

\subsection{The Density Fluctuation Parameter in \HII Regions}

In practice, $G_{\rm scatt}$ and $\tilde{F}$ are only constrained when the distances of both the pulsar and scattering layer are known. There are five pulsars in our sample of scattering measurements that have precise parallax distances, J0357$+$5236, J0358$+$5413, J0601$-$0527, J1614$-$2230, and J1643$-$1224. The first two lie towards the Galactic anticenter and the latter three lie at Galactic latitudes $|b|>10^\circ$. We estimate the DM contributions of the \HII regions along these LOSs by assuming the best-fit plane-parallel model for the electron density structure of the thick disk, which has a mid-plane density of $n_0 = 0.015 \pm 0.001$ cm$^{-3}$ and a scale height of $1.57\pm0.15$ kpc \citep{ocker2020}. While this model is most accurate for high Galactic latitudes, it appears to give reasonable estimates for the two pulsars in the plane, J0357$+$5236 and J0358$+$5413, for reasons discussed below.

The DM contributions of the ISM and \HII regions for these pulsars, based on the \cite{ocker2020} plane-parallel model, are shown in Table~\ref{tab:Ftilde}, along with resulting estimates of $\tilde{F} G_{\rm scatt}$ that assume the observed scattering is dominated by the \HII region. Pulsars J0357$+$5236 and J0358$+$5413 lie in the plane at Galactic longitudes $l \approx 149^\circ$, and both intersect the \HII region S205 \citep{mitra2003}. The smooth plane-parallel model tends to be inaccurate at low Galactic latitudes, due primarily to the presence of spiral arms and increased density variance associated with discrete structures. In this case, the source of the density variance is known, and assuming the mid-plane density of $0.015$ cm$^{-3}$ yields DM contributions from the \HII region that are very similar for both pulsars, and comparable to the \HII region DMs estimated for the pulsars at higher latitudes.

\begin{deluxetable*}{c C C C C C}\label{tab:Ftilde}
\tablewidth{1.5\textwidth}
\tablecaption{\HII Region Density Fluctuation Parameter for Pulsars with Parallax Distances}
\tablehead{\colhead{PSR} & \colhead{\hspace{0.5cm}$\tau$}\hspace{0.5cm} & \colhead{\hspace{0.5cm}$\rm DM_{ISM}$}\hspace{0.5cm} & \colhead{\hspace{0.5cm}$\rm DM_{HII}$}\hspace{0.5cm} & \colhead{\hspace{0.5cm}$\tilde{F}G_{\rm scatt}/10^{-3}$}\hspace{0.5cm} & \colhead{\hspace{0.5cm}$\tilde{F}/10^{-4}$}\hspace{0.5cm}  \\ \colhead{} & \colhead{\hspace{0.5cm}(ms at 1 GHz)}\hspace{0.5cm} & \colhead{\hspace{0.5cm}(pc cm$^{-3}$)}\hspace{0.5cm} & \colhead{\hspace{0.5cm}(pc cm$^{-3}$)}\hspace{0.5cm} & \colhead{\hspace{0.5cm}(pc$^2$ km)$^{-1/3}$}\hspace{0.5cm} & \colhead{\hspace{0.5cm}(pc$^2$ km)$^{-1/3}$}}\hspace{0.5cm}
\startdata
J0357$+$5236  & 2.8\times10^{-3} & 49\pm8 & 55\pm8 & 5\pm1 & 6.1^{+4.4}_{-2.0} \\
J0358$+$5413  & 3.5\times10^{-4} & 16\pm3 & 41\pm3 & 2.3\pm0.4 & 3.7^{+5.1}_{-1.5} \\
J0601$-$0527  & 4.5\times10^{-3} & 27\pm3 & 54\pm3 & 15\pm2 & 6.4^{+1.6}_{-1.5} \\
J1614$-$2230  & 1.6\times10^{-4} & 11\pm1 & 24\pm1 & 2.9\pm0.4 & 4.0^{+1.4}_{-1.3} \\
J1643$-$1224  & 4.9\times10^{-2} & 12\pm1 & 50\pm1 & 270\pm40 & 370^{+130}_{-100} \\
\enddata
\tablecomments{Values of $\rm DM_{ISM}$ correspond to the DM contribution of the thick disk based on the best-fit plane-parallel model of \cite{ocker2020}, and the \HII region contributions are estimated as $\rm DM_{HII} = DM_{tot} - DM_{ISM}$. Errors correspond to the $15\%$ and $85\%$ confidence intervals. Errors on $\tilde{F} G_{\rm scatt}$ include the uncertainty in $\rm DM_{HII}$ and an assumed $10\%$ uncertainty in $\tau$ {(although estimates of $\tau$ could in principle vary by more than $10\%$ due to estimation errors and refraction at larger scales; e.g. \citealt{ramachandran2006})}. Errors on $\tilde{F}$ further include uncertainties in the distances of the pulsars and \HII regions, which yield uncertainties on the LOS path lengths through the \HII regions; these estimates assume that the entire path through the \HII region contributes to scattering, which may not be the case. The fluctuation parameter conventionally used to estimate scattering measures, $F_c = \tilde{F}l_{\rm i}^{1/3}$, would be $10\times \tilde{F}$ for $l_{\rm i} = 10^3$ km.}
\end{deluxetable*}

The density fluctuation parameter $\tilde{F}$ is estimated by assuming normal PDFs for the pulsar parallaxes, \HII region distances, DMs, and $\tau$. A uniform PDF is evaluated for $L$ that is based on the path length through a sphere with radius given by the angular diameter of the \HII region and its distance, along with the pulsar's transverse distance from the center of the \HII region. The PDF width for $L$ is determined by the uncertainty in the \HII region distance, except in the case of J0601$-$0527 where the \HII region distance is a lower limit and we instead evaluate the PDF for the minimum physical diameter that allows the pulsar LOS to have a non-zero path length through the \HII region. Assuming that all of the observed scattering is attributable to the \HII regions yields the estimates of $\tilde{F}$ shown in Table~\ref{tab:Ftilde}, which can be converted for an assumed inner scale to $\tilde{F}l_{\rm i}^{1/3} \equiv F_c$, the fluctuation parameter that is conventionally used to estimate scattering measures. 

We find $\tilde{F}$ ranges from $\sim 10^{-4} - 10^{-2}$ (pc$^2$ km)$^{-1/3}$, with the largest value corresponding to J1643$-$1224. This larger value is comparable to expectations for $l_{\rm o} \sim 20$ pc and $l_{\rm i} \sim 100$ km, for $\zeta \approx \epsilon^2 \approx f \sim 1$. Leaving $l_{\rm o}$, $l_{\rm i}$, $\zeta$, and $f$ at the same nominal values as in Equation~\ref{eq:ftilde} would require $\epsilon \ll 1$ to explain the small values of $\tilde{F}$ found for the other \HII regions in Table~\ref{tab:Ftilde}. This statement is agnostic to changes in $\zeta$ and $f$ because $f \leq 1$ and $\zeta \geq 1$.  

The DM and scattering contributions of \HII regions S7 and S27 were previously assessed for J1614$-$2230 and J1643$-$1224 by \cite{ocker2020}.  Here we use an updated parallax distance for J1643$-$1224 from \cite{ding2023}, and unlike \cite{ocker2020} we do not subtract an estimate of the ISM contribution to $\tau$, nor do we assume the \HII region diameter is equivalent to the path length through the region. These differences yield values of $F_c$ that are smaller than those reported by \cite{ocker2020} for these two pulsars. 

The fluctuation parameters shown in Table~\ref{tab:Ftilde} are smaller than values that would be inferred if we instead assumed that the pulsar scattering occurs in the extended ISM (in which case $G_{\rm scatt} = 1/3$). We have also assumed that the pulsar's entire path length through the \HII region contributes to the observed scattering. If the scattering is confined instead to the \HII region boundary, or to an overdensity inside the \HII region, then $\tilde{F}$ will tend to be larger; e.g., re-evaluating the scattering time of J1643$-$1224 for a screen width $L\sim1$ pc and a DM contribution $\sim 1$ pc cm$^{-3}$ yields $G_{\rm scatt} \approx 300$ and $\tilde{F} \sim 4$ (pc$^2$ km)$^{-1/3}$. While these degeneracies between screen width, DM, and $\tilde{F}$ make it difficult to further interpret the values of $\tilde{F}$ shown in Table~\ref{tab:Ftilde}, we can nonetheless assess the degree to which variations in DM, $G_{\rm scatt}$, or $\tilde{F}$ dominate the extreme differences in $\tau$ between nearby pulsars intersecting \HII regions (such as those with parallax distances), and more distant pulsars intersecting \HII regions, which show the largest measured scattering. 

\begin{figure}
    \centering
    \includegraphics[width=0.45\textwidth]{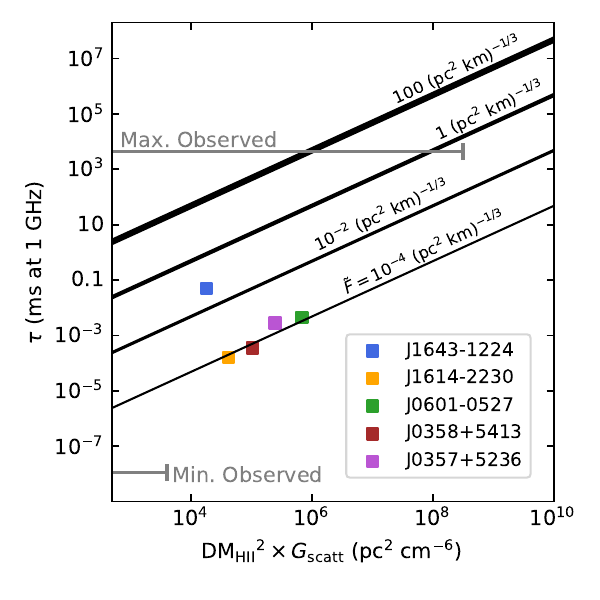}
    \caption{Scattering delay $\tau$ in ms at 1 GHz vs. ${\rm DM_{HII}}^2\times G_{\rm scatt}$, for four different values of the density fluctuation parameter $\tilde{F}$ (black solid lines). Colored squares indicate the inferred values of ${\rm DM_{HII}}^2\times G_{\rm scatt}$ for the five pulsars with parallax distances that intersect \HII regions, which yield the values of $\tilde{F}$ shown in Table~\ref{tab:Ftilde} if the \HII regions dominate the observed scattering. {The largest and smallest published scattering delays are shown by the grey horizontal lines, corresponding to PSRs J1550$-$5418 ($\tau_{\rm1 GHz} \approx 4000$ ms) and J0953$+$0755 ($\tau_{\rm1 GHz} \approx 1\times10^{-8}$ ms). For these two extremes in $\tau$ we show conservative upper limits on ${\rm DM^2} G_{\rm scatt}$ that assume the total observed DM comes from a scattering screen halfway between the source and observer, and a screen width $\sim 0.1\%$ of the distance to the source. J1550$-$5418 intersects at least one \HII region, and its observed scattering can be explained for reasonable values of ${\rm DM_{HII}}^2\times G_{\rm scatt}$ if $\tilde{F}$ is $\gtrsim1$ (pc$^2$ km)$^{-1/3}$.}}
    \label{fig:tau_vs_dmG}
\end{figure}

We focus on comparing two extremes in the distribution of $\tau$ for pulsars intersecting \HII regions, {illustrated in Figure~\ref{fig:tau_vs_dmG}}: $\tau \sim 10^{-4}$ ms and $\tau \sim 10^2$ ms at 1 GHz. In the low-$\tau$ case, we have found fiducial values of $\tilde{F} \sim 10^{-4} - 10^{-2}$ (pc$^2$ km)$^{-1/3}$. Assuming these values of $\tilde{F}$ apply to the high-$\tau$ case yields ${\rm DM}^2\times G_{\rm scatt} \sim 10^{8} - 10^{10}$ pc cm$^{-3}$. Such high values of ${\rm DM}^2\times G_{\rm scatt}$ would require $L < 1$ pc and $\rm DM > 10^3$ pc cm$^{-3}$, corresponding not only to implausibly high electron densities {but also DMs larger than typically observed}. {This discrepancy could potentially be alleviated in an alternative model where scattering arises from filaments or sheets that are highly inclined relative to the LOS \citep{pen2014}.} The two extremes in $\tau$ could conversely be reconciled for reasonable values of ${\rm DM}^2\times G_{\rm scatt}$ if $\tilde{F}$ is much larger ($\sim 1 - 100$ (pc$^2$ km)$^{-1/3}$) for the high-$\tau$ LOSs.

If the scattering of these pulsars is dominated by their foreground \HII regions, then {application of the ionized cloudlet model leaves} two possible conclusions: either the distribution of $\tilde{F}$ for \HII regions varies dramatically (perhaps due to underlying properties of the \HII regions, such as age), or $\tilde{F}$ is roughly the same for all \HII regions and it is variations in ${\rm DM}^2\times G_{\rm scatt}$ that lead to the large range in $\tau$. We consider the latter conclusion less likely because large values of ${\rm DM}^2\times G_{\rm scatt}$ would require {that \HII regions contribute $\rm DM > 10^3$ pc cm$^{-3}$, which combined with the ISM would imply DMs greater than observed.} A larger sample of pulsars with $\rm DM > 100$ pc cm$^{-3}$ and known distances could be used to search for correlations between inferred estimates of $\tilde{F}$ and other \HII region properties, which may reveal whether variations in $\tilde{F}$ between different \HII regions can be plausibly linked to other tracers of their internal properties.

A complementary test of radio scattering by \HII regions is through Fourier analysis of pulsar scintillation (i.e., scintillation arcs; \citealt{stinebring2001}), which constrains the locations of scattering material and disentangles scattering from multiple locations along the LOS. Two of the pulsars in our sample, J0358$+$5236 and J1643$-$1224, have scintillation arcs that are consistent with scattering locations in their foreground \HII regions, affirming that the \HII regions contribute to the observed scattering \citep{mall2022,ocker2024_scintarcs}. However, in both of these cases, multiple arcs are detected at varying distances from the observer: for J0358$+$5236, at least three other scattering screens are located beyond the \HII region \citep{ocker2024_scintarcs}, and for J1643$-$1224 there may be one additional screen beyond the \HII region \citep{mall2022}, {albeit with less scattering strength}. In principle, such observations could be used to constrain the amount of scattering contributed by each screen along the LOS, including the specific scattering contributions of the \HII regions, but such studies are complicated by a variety of effects that modify the observed scintillation, including but not limited to radio-frequency dependence (typically one or two screens dominate scintillation at low frequencies, whereas many more screens contribute at higher frequencies). Scintillation observations of pulsars at distances $<400$ pc suggest that the DM contributions of scattering screens associated with scintillation arcs are extremely small, because many screens are detected for pulsars with total DMs $<5$ pc cm$^{-3}$ (\citealt{ocker2024_scintarcs}; D. Reardon et al., in review). These results may support a scenario in which variations in $\tilde{F}$ (and $G_{\rm scatt}$) dominate variations in $\tau$, although the degree to which results from pulsar scintillation arcs are applicable to more distant pulsars in the high-DM, high-$\tau$ regime remains unclear, given that scintillation arcs tend to be observable only for nearby, lower DM pulsars{, due in part to observational constraints that limit scintillation arc detectability to bright pulsars that can be observed with sufficient frequency resolution} \citep{stinebring2022,main2023}. 

\section{Discussion}\label{sec:implications}

\begin{figure}
    \centering
    \includegraphics[width=0.47\textwidth]{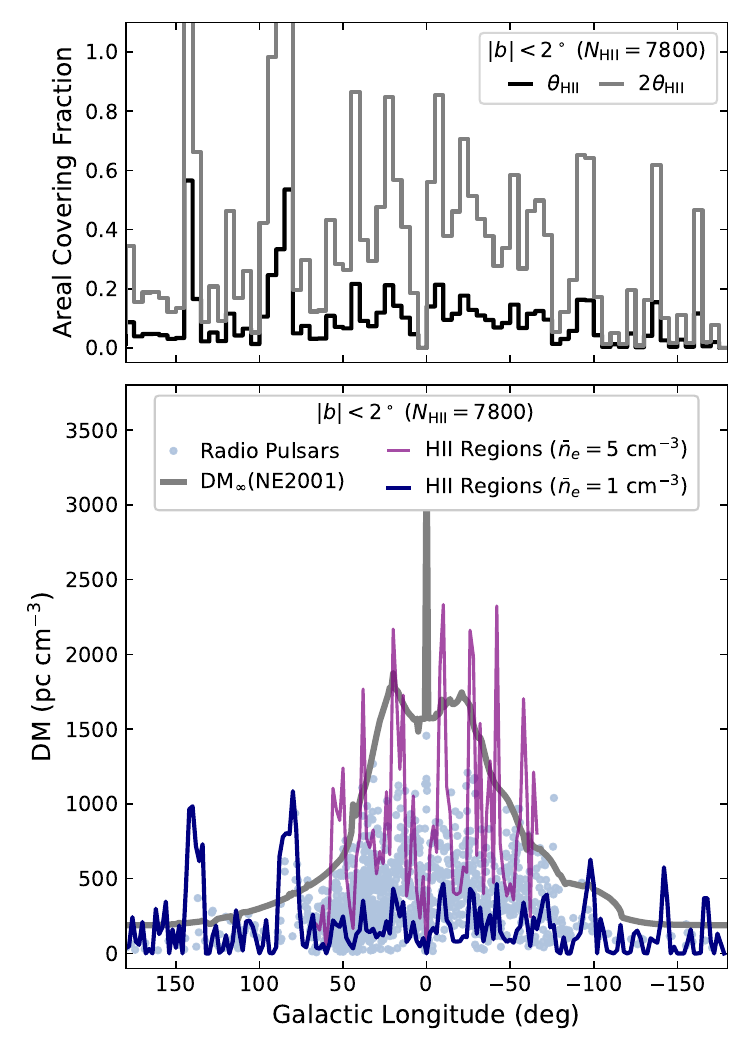}
    \caption{{\textit{Top:} Areal covering fraction vs. Galactic longitude of $\approx 7800$ \HII regions in the WISE Catalog at Galactic latitudes $|b|<2^\circ$. The areal covering fraction is evaluated in $5^\circ$ wide longitude bins using the angular radii given by the WISE Catalog ($\theta_{\rm HII}$; black line) and twice the angular radii ($2\theta_{\rm HII}$; grey line). \textit{Bottom:} Simulation of the maximum DM contribution vs. Galactic longitude from all \HII regions in the WISE Catalog at $|b|<2^\circ$. The dark blue curve gives the maximum possible DM for LOSs passing within $2\theta_{\rm HII}$ of \HII regions, assuming that \HII regions are described by a mean electron density $\bar{n}_e = 1$ cm$^{-3}$. The purple curve shows the maximum possible DM from \HII regions assuming $\bar{n}_e = 5$ cm$^{-3}$, restricted to longitudes $|l|<70^\circ$ to illustrate how with larger densities, \HII regions could account for a significant fraction of DMs in the inner Galaxy. The distribution of DMs for radio pulsars at $|b|<2^\circ$ is shown by the light blue points, and the maximum DM predicted by NE2001 at $b=0^\circ$ is shown by the solid grey curve.}}
    \label{fig:covering_fraction}
\end{figure}

{Our study has been limited thus far to comparing the pulsar population to \HII regions with distances, which only constitute $\approx 1/3$ of the $\approx 8400$ sources in the WISE Catalog {(\S~\ref{sec:catalogs})}. To illustrate the potential impact of the full \HII region population on pulsar observables, we estimate both the areal covering fraction of \HII regions and the maximum possible DM contribution of \HII regions as functions of Galactic longitude. The top panel of Figure~\ref{fig:covering_fraction} shows the areal covering fraction of all regions in the WISE Catalog at Galactic latitudes $|b|<2^\circ$, based on the angular radii ($\theta_{\rm HII}$) given in the catalog. Using the cataloged $\theta_{\rm HII}$, we find an areal covering fraction that is $\approx 20\%$ across much of the inner Galaxy. Using the larger angular extent ($2\theta_{\rm HII}$) assumed in our intersection analysis yields an areal covering fraction $\approx 50\%$ across the inner Galaxy, although the covering fraction varies substantially with longitude and in some cases is $>80\%$ (notably rising to $>100\%$ around the Cygnus region). The bottom panel of Figure~\ref{fig:covering_fraction} shows simulated estimates of the maximum possible DM from \HII regions for extragalactic LOSs at $|b|<2^\circ$. We estimate the maximum DM by sampling $5000$ LOSs uniformly distributed over all Galactic longitudes and identifying the \HII regions in the WISE Catalog that every LOS intersects within $2\theta_{\rm HII}$. Each \HII region's distance is randomly drawn from a uniform distribution spanning 1 to 10 kpc, and the LOS path length through {the} \HII region is evaluated using the LOS impact parameter. Figure~\ref{fig:covering_fraction} shows the DM integrated through all intervening \HII regions for two mean electron densities describing {their interiors}: 1 cm$^{-3}$ and 5 cm$^{-3}$. For illustrative purposes, the maximum DM distribution is binned by every $2^\circ$ in longitude. This simulation does not include the ISM, which is why the DM distribution shown in Figure~\ref{fig:covering_fraction} drops to $0$ pc cm$^{-3}$ for some LOSs. When compared to the observed DMs of Galactic pulsars and the maximum DM predicted by NE2001, this simple simulation suggests that even for a modest electron density (1 cm$^{-3}$), \HII regions could account for $20-25\%$ of the total DM in the Galactic plane for longitudes $|l|\lesssim70^\circ$, and that for $|l|>70^\circ$ \HII regions can contribute the majority of DM. Allowing for larger interior densities would imply that \HII regions account for a much higher fraction ($\gtrsim50\%$) of the total DM.}

{Our results suggest that \HII regions form a significant constituent of the electron density content probed by the pulsar population. Accounting for \HII regions is thus critical to understanding present-day uncertainties in Galactic electron density models, which are the primary tool for estimating distances to both pulsars and fast radio bursts (FRBs) that lack independent distance measures. Galactic electron density models are fundamental to a broad range of studies; within pulsar astronomy alone, the distances inferred from these models are used to characterize the pulsar velocity distribution \citep{arzoumanian2002} and underlying dynamics at birth \citep{hansen1997}, estimate the formation rate of double neutron star systems \citep{narayan1991}, constrain the Shklovskii effect in pulsar timing \citep{1970SvA....13..562S}, and test perturbations to General Relativity \citep{kramer2006}. These examples are only a handful of the purposes for which Galactic electron density models have been used, and the model uncertainties have correspondingly broad implications.}

\HII regions should be incorporated into a next-generation Galactic electron density model. While early electron density models of the local ISM inferred pulsar distances based on a statistical distribution of \HII regions \citep{prentice-terhaar1969}, modern electron density models based on pulsars do not explicitly include the known \HII region population, which is effectively subsumed into spiral arm components. Evidence for the electron density structure of the spiral arms is inferred from a spatially-dependent deficit in the DM predicted by smooth models that lack spiral arms \citep{1993ApJ...411..674T,gbc01}. Spiral arms are included in models such as NE2001 and YMW16 as smooth overdensities, and to account for inhomogeneities in the arms due to, e.g., star-forming regions, \HII regions, supershells, and supernova remnants, NE2001 includes LOS-specific clumps and voids that are required to explain certain DM and scattering measurements. Some of the clumps used in NE2001 include \HII regions identified in our analysis. YMW16 does not include specific clumps or voids whatsoever. Most of these prior models suggested that \HII regions are not statistically significant contributors to the DM and scattering of the observed pulsar population. 

Our findings leave a potential dilemma for future electron density models. \HII regions could be included on a case-by-case basis, as done for TC93 and NE2001, or a spiral arm model could directly mimic the known \HII region population. Both of these options would be cumbersome to employ. The former would require calibrating the DM and scattering contributions of up to 100s of \HII regions on a pulsar-by-pulsar basis. The latter option would likely involve a `universal' \HII region model that accounts for a range of \HII region sizes and interior densities, which would need to be calibrated against 100s of pulsars with known distances intersecting \HII regions at a range of impact parameters. {These approaches are under consideration}, and \HII regions with the most extreme DM and scattering contributions will be added on a case-by-case basis to an in progress, extended version of NE2001p \citep{ocker2024_ne2001p}.

Regardless of how future electron density models are constructed, the known \HII region population can be used to constrain uncertainties in the currently used models. In general, LOSs with foreground \HII regions will tend to have distances that are overestimated by NE2001 and YMW16 {(\S\ref{sec:intersections}). However, the tendency to over- or under-estimate distances will be spatially dependent. {As shown in Figure~\ref{fig:spiral_arms},} in the second and third Galactic quadrants NE2001 places spiral arms at larger distances from the Sun than the \cite{reid2019} arms, suggesting that pulsars in these regions will have overestimated distances. Conversely, in the third and fourth Galactic quadrants NE2001 places portions of the Carina-Sagittarius and Crux-Scutum arms closer to the Sun than \cite{reid2019}, suggesting that pulsars with LOSs tangent to those arms will tend to have underestimated distances in those quadrants. While the spiral arm locations of both NE2001 and \cite{reid2019} have substantial uncertainties, the \HII region locations in Figure~\ref{fig:spiral_arms} are more consistent with the \cite{reid2019} arms. All log-periodic spiral arm models are idealized, approximate views of the Milky Way's global structure, and the exact number of primary arms and roles of kinks and bifurcations remain subject to disagreement \citep[e.g.][]{xu2023,denyshchenko2024}. Such problems would be largely alleviated if the observed pulsar population reaches the critical mass needed to directly perform tomography of the ionized ISM. 

In addition to their use as distance estimators, Galactic models are critical to constraining the FRB Galactic foreground, which induces noise in FRB-based inferences of both host galaxy and intergalactic medium (IGM) properties \citep{petroff19,2019ARA&A..57..417C}. While {some} FRB surveys have prioritized high Galactic latitudes, where \HII regions are rare, current and future surveys {(including the largest thus far, CHIME; \citealt{chimecat1})} inevitably cover low latitudes where uncertainties in Galactic models exacerbate the identification of local universe FRBs with ambiguously small DMs \citep{ravi2023}, {which also provide the most stringent constraints on FRB source models \citep[e.g.][]{bhardwaj2021,kirsten2022}.} Simultaneous searches for both new pulsars and radio bursts from Galactic magnetars are also most successful in the plane \citep{bochenek2020,chime2020,good2021,dong2023_chimepulsar}, and these latitudes will be increasingly important as surveys like DSA-2000 \citep{hallinan2019,ravi2019} and CHORD \citep{vanderlinde2019_chord} expand the sensitivity phase space for pulsar, magnetar, and FRB discovery. Figure~\ref{fig:dm_vs_lon} {roughly indicates} the degree of `noise' that \HII regions contribute to the Galactic DM foreground as a function of Galactic longitude; it is already apparent from this figure that there are certain regions of the sky where FRB surveys are more likely to misidentify Galactic sources as extragalactic, because Galactic electron density models are used to distinguish between Galactic and extragalactic FRB candidates. 

\HII regions are also important for interpreting FRB host galaxy contributions to DM. When the host galaxy inclination angle is large (edge-on), then \HII regions can contribute 10s to 100s pc cm$^{-3}$ to the host DM, depending on the FRB location. H$\alpha$ EMs are frequently used as an alternative predictor of the host DM contribution, even though these EMs are based on single-slit measurements that integrate over the whole galaxy \citep{2017ApJ...834L...7T,ocker2022}. Understanding the degree to which \HII regions vs. the diffuse ISM contribute to the observed H$\alpha$ intensity may be especially relevant when the host DM appears to be large (or small) relative to the observed H$\alpha$ emission, because it is typically unclear whether the FRB's circumsource medium (e.g., a supernova remnant or wind nebula) contributes significantly to the host DM \citep{ocker2022,caleb2022}. Similarly, constraints on the density fluctuation parameter $\tilde{F}$ in host galaxies typically assume that the FRB scattering arises from the host's extended ISM \citep{cordes2022,ocker2022,cassanelli2023,caleb2023}, but for high host inclination angles, one or a few \HII regions could dominate the scattering contribution of the host.   

\section{Summary}\label{sec:summary}

We have shown that the majority of pulsars with $\rm DM > 600$\,pc\,cm$^{-3}$ and $\tau > 10$\,ms at~1\,GHz intersect \HII regions with distances in the WISE and HH14 catalogs, and that nearly all pulsars with DMs greater than the maximum predictions of NE2001 and YMW16 lie behind \HII regions. For a subset of about 500 pulsars with scattering measurements, we have individually assessed the validity of identified intersections and found a false positive rate of about $20\%$ and a false negative rate $\leq 30\%$. A catalog of the pulsar and \HII region pairs is provided in Appendix~\ref{app:h2list}; these pulsars would be prime follow-up targets for Faraday RM studies. Intersections were likely missed because they were identified using only \HII regions with published distances, and because of the assumed impact parameter cutoff. Combined with the number of validated intersections, the false positive and negative rates suggest that as much as a third of the known pulsar population may intersect \HII regions. {The fraction of the full pulsar population impacted by foreground \HII regions will almost certainly be larger, because current pulsar surveys are biased towards nearby pulsars, and \HII region intersections will increase for sightlines deeper into the Galactic plane.} Due to selection effects that bias the intersection sample towards LOSs through the inner Galaxy, we anticipate that more intersections for pulsar DMs $<600$ pc cm$^{-3}$ will be found as the sample of both pulsar and \HII region distances increases. A detailed examination of pulsar LOSs and \HII regions covered by radio recombination maps suggests that \HII regions contribute 10s to 100s of pc cm$^{-3}$ in electron column, depending on the LOS. Our assessment of pulsar scattering measurements suggests that if \HII regions dominate the scattering of background pulsars, then the density fluctuations parameterized by $\tilde{F}$ must vary significantly between \HII regions. {Simulating the full \HII region population assuming a modest interior electron density suggests that at low latitudes, \HII regions can account for $\gtrsim20\%$ of the total DM contribution of the Galactic disk, and that for some LOSs (e.g. near the Cygnus region), \HII regions contribute the majority of free electrons along the LOS.}

These results can be directly tested by cross-correlating pulsar LOSs spanning a range of impact parameters through \HII regions with detailed spectroscopic information, which will constrain any relation between \HII region properties and the turbulent density fluctuations probed by pulsar scattering, in addition to yielding the mean electron density profiles of \HII regions around different ionizing star types. Such science cases may soon be achievable with the SDSS-V Local Volume Mapper \citep{drory2024_lvm}.

\acknowledgements{SKO is supported by the Brinson Foundation through the Brinson Prize Fellowship Program. SKO, JL, and JMC are members of the NANOGrav Physics Frontiers Center (NSF award PHY-2020265). SKO is grateful to Liam Connor, Casey Law, Kritti Sharma, Jakob Faber, Myles Sherman, Nikita Gosogorov, Nick Konidaris, and Francesco Iraci for fruitful discussions, to the anonymous reviewer for their feedback, and to Jack Madden for lending his eyes to Figure 5. Caltech and Carnegie Observatories are located on the traditional and unceded lands of the Tongva people.}

\bibliography{master_bib}

\appendix
\restartappendixnumbering

\section{Catalog of Pulsars Intersecting \HII Regions}\label{app:h2list}

Table~\ref{tab:intersections1} gives the full list of pulsars with scattering measurements that are identified as intersecting \HII regions based on our selection criteria.

\startlongtable
\begin{deluxetable*}{L C C C C C c | C C C C C C}\label{tab:intersections1}
\tabletypesize{\scriptsize}
\tablecaption{Pulsars Intersecting \HII Regions}
\tablehead{\multicolumn{7}{c}{Pulsar} \vline & \multicolumn{5}{c}{\HII Region} \\ \hline
            \colhead{Name} & \colhead{$l$} & \colhead{$b$} & \colhead{DM} & \colhead{$\tau$} & 
            \colhead{$D$} & \colhead{Method} \vline & \colhead{$l$} & \colhead{$b$} & \colhead{$D$} & \colhead{Ref.} & \colhead{Notes} \\
            \colhead{} & \colhead{$(^{\circ})$} & \colhead{$(^{\circ})$} & \colhead{(pc cm$^{-3}$)} &
            \colhead{(ms at 1~GHz)} & \colhead{(kpc)} & \colhead{} \vline & \colhead{$(^{\circ})$} & \colhead{$(^{\circ})$} & \colhead{(kpc)} & \colhead{} & \colhead{} }

\startdata
\rm J0357+5236 & 149.099 & -0.522 & 103.7 & 2.8\times10^{-3} & 3.33\pm0.72 & P & 148.540 & -0.240 & 0.85\pm0.15 & \rm HH14 & \rm S205\\
\rm J0358+5413 & 148.190 & 0.811 & 57.1 & 3.5\times10^{-4} & 1.10\pm0.20 & P & 148.540 & -0.240 & 0.85\pm0.15 & \rm HH14 & \rm S205\\
\rm J0601-0527 & 212.199 & -13.481 & 80.5 & 4.5\times10^{-3} & 2.08\pm0.17 & P & 213.704 & -12.606 & >0.61 & \rm HH14 & \\
\rm J0837-4135 & 260.904 & -0.336 & 147.2 & 1.0\times10^{-2} & 1.50\pm0.45 & Y & 261.381 & 0.841 & 0.67\pm0.17 & \rm HH14 & \\
\rm J0857-4424 & 265.457 & 0.822 & 184.4 & 0.11 & 2.83\pm0.85 & Y & 265.151 & 1.454 & 0.91\pm0.26 & \rm HH14 & \\
\rm J1046-5813 & 287.065 & 0.733 & 240.2 & 0.13 & 4.37\pm1.31 & N & 287.247 & 0.355 & 2.50\pm0.10 & \rm WISE & \rm RCW52\\
\rm J1056-6258 & 290.292 & -2.966 & 320.6 & 0.45 & 2.99\pm0.90 & N & 290.323 & -2.983 & [0.93,4.74] & \rm WISE & \\
\rm J1114-6100 & 291.443 & -0.321 & 676.7 & 21 & 15.48\pm4.64 & N & 291.467 & -0.142 & 7.20\pm1.20 & \rm WISE & \rm multiple\\
\rm J1133-6250 & 294.213 & -1.296 & 567.8 & 22 & 11.94\pm3.58 & N & 294.793 & -1.329 & 3.60\pm0.60 & \rm WISE & \rm multiple\\
\rm J1138-6207 & 294.506 & -0.463 & 518.9 & 26 & 9.56\pm2.87 & N & 294.453 & -0.520 & [0.58,6.20] & \rm WISE & \\
\rm J1305-6203 & 304.561 & 0.772 & 468.8 & 10 & 11.59\pm3.48 & Y & 304.583 & 0.582 & >2.17 & \rm HH14 & \\
\rm J1316-6232 & 305.848 & 0.191 & 966.4 & 1.0\times10^{3} & 25.00\pm7.50 & Y & 305.789 & 0.138 & 5.00\pm0.10 & \rm WISE & \rm multiple\\
\rm J1327-6222 & 307.074 & 0.204 & 318.5 & 0.98 & 5.48\pm1.64 & N & 308.653 & -0.510 & >0.50 & \rm HH14 & \\
\rm J1338-6204 & 308.372 & 0.305 & 640.3 & 28 & 12.36\pm3.71 & Y & 308.747 & 0.547 & 5.30\pm0.80 & \rm WISE & \\
\rm J1341-6220 & 308.730 & -0.035 & 719.6 & 26 & 12.60\pm3.78 & Y & 309.075 & 0.172 & 5.40\pm1.20 & \rm WISE & \\
\rm J1349-6130 & 309.813 & 0.587 & 283.9 & 6.4 & 5.48\pm1.64 & Y & 309.905 & 0.373 & 5.50\pm0.10 & \rm WISE & \\
\rm J1359-6038 & 311.239 & 1.126 & 293.7 & 0.47 & 5.15\pm1.54 & N & 312.305 & 0.663 & >0.94 & \rm HH14 & \\
\rm J1406-6121 & 311.841 & 0.203 & 537.3 & 48 & 8.09\pm2.43 & N & 311.893 & 0.087 & 5.70\pm2.40 & \rm WISE & \rm multiple\\
\rm J1412-6145 & 312.324 & -0.366 & 512.5 & 19 & 7.79\pm2.34 & N & 311.850 & -0.540 & 1.50\pm0.40 & \rm HH14 & \rm RCW83\\
\rm J1413-6141 & 312.462 & -0.337 & 667.6 & 42 & 10.01\pm3.00 & N & 311.850 & -0.540 & 1.50\pm0.40 & \rm HH14 & \rm RCW83\\
\rm J1511-5835 & 320.289 & -0.508 & 329.4 & 23 & 5.12\pm1.54 & Y & 320.109 & -0.510 & >0.93 & \rm HH14 & \rm multiple\\
\rm J1512-5759 & 320.772 & -0.108 & 627.5 & 5.5 & 7.35\pm2.20 & N & 320.893 & -0.410 & >2.97 & \rm HH14 & \rm multiple\\
\rm J1514-5925 & 320.284 & -1.482 & 192.5 & 16 & 3.91\pm1.17 & Y & 320.109 & -0.510 & >0.93 & \rm HH14 & \rm multiple\\
\rm J1527-5552 & 323.638 & 0.590 & 370.1 & 0.43 & 5.38\pm1.61 & Y & 323.680 & 0.630 & [3.00,10.20] & \rm WISE & \\
\rm J1543-5459 & 326.025 & -0.044 & 345.0 & 36 & 4.82\pm1.45 & N & 326.474 & -0.290 & [3.54,10.10] & \rm WISE & \\
\rm J1550-5418 & 327.237 & -0.132 & 830.0 & 4.3\times10^{3} & 9.54\pm2.86 & N & 326.985 & -0.158 & >3.37 & \rm HH14 & \rm multiple\\
\rm J1551-5310 & 328.033 & 0.669 & 491.6 & 1.3\times10^{2} & 7.51\pm2.25 & N & 328.117 & 0.570 & 5.34\pm0.40 & \rm WISE & \\
\rm J1600-3053 & 344.090 & 16.451 & 52.3 & 1.2\times10^{-2} & 1.89\pm0.57 & Y & 347.220 & 20.240 & 0.20\pm0.40 & \rm HH14 & \rm S1\\
\rm J1614-2230 & 352.636 & 20.192 & 34.5 & 1.6\times10^{-4} & 0.77\pm0.05 & P & 349.840 & 22.260 & 0.20\pm0.04 & \rm HH14 & \rm S7\\
\rm J1614-5048 & 332.206 & 0.172 & 582.4 & 35 & 7.93\pm2.38 & N & 332.275 & -0.003 & [3.10,11.40] & \rm WISE & \\
\rm J1640-4715 & 337.713 & -0.439 & 586.3 & 44 & 6.43\pm1.93 & N & 337.684 & -0.342 & [2.90,12.20] & \rm WISE & \\
\rm J1643-1224 & 5.669 & 21.218 & 62.4 & 4.9\times10^{-2} & 0.91\pm0.08 & P & 6.280 & 23.580 & 0.20\pm0.04 & \rm HH14 & \rm S27\\
\rm J1644-4559 & 339.193 & -0.195 & 478.7 & 11 & 5.09\pm1.53 & N & 339.134 & -0.376 & 3.00\pm0.40 & \rm WISE & \\
\rm J1707-4053 & 345.718 & -0.197 & 351.8 & 46 & 4.46\pm1.34 & N & 345.495 & 0.326 & 2.00\pm0.25 & \rm HH14 & \\
\rm J1715-3859 & 348.194 & -0.348 & 806.2 & 2.2\times10^{2} & 9.42\pm2.83 & N & 348.691 & -0.825 & 3.40\pm0.30^\dagger & \rm WISE & \rm RCW122\\
\rm J1717-3425 & 352.120 & 2.025 & 583.5 & 16 & 25.00\pm7.50 & Y & 352.440 & 2.260 & 1.00\pm0.20 & \rm HH14 & \rm S10\\
\rm J1717-3737 & 349.491 & 0.182 & 522.7 & 35 & 5.77\pm1.73 & N & 350.000 & 0.200 & 1.60\pm0.50 & \rm HH14 & \rm RCW125\\
\rm J1720-3659 & 350.332 & 0.100 & 379.0 & 12 & 4.65\pm1.39 & N & 350.000 & 0.200 & 1.60\pm0.50 & \rm HH14 & \rm RCW125\\
\rm J1721-3532 & 351.687 & 0.670 & 496.8 & 94 & 5.76\pm1.73 & N & 351.662 & 0.518 & 1.35\pm0.15^\dagger & \rm HH14 & \rm RCW128\\
\rm J1722-3632 & 350.934 & -0.001 & 416.2 & 9.8 & 4.44\pm1.33 & N & 351.040 & 0.660 & 1.35\pm0.15^\dagger & \rm HH14 & \rm S8\\
\rm J1723-3659 & 350.682 & -0.409 & 254.4 & 1.9 & 3.62\pm1.09 & N & 350.991 & -0.530 & [2.90,13.20] & \rm WISE & \\
\rm J1731-3123 & 356.233 & 1.354 & 354.5 & 15 & 7.29\pm2.19 & Y & 355.890 & 1.610 & 1.20\pm0.25 & \rm HH14 & \rm S13\\
\rm J1745-2900 & 359.944 & -0.047 & 1778.0 & 1.3\times10^{3} & 8.49\pm2.54 & N & 359.964 & -0.100 & 5.52\pm1.05^\dagger & \rm WISE & \rm multiple\\
\rm J1745-2912 & 359.788 & -0.175 & 1130.0 & 3.6\times10^{3} & 8.41\pm2.52 & N & 359.757 & -0.351 & 2.67\pm0.16^\dagger & \rm WISE & \rm multiple\\
\rm J1746-2849 & 0.134 & -0.030 & 1456.0 & 1.5\times10^{3} & 8.47\pm2.54 & N & 0.320 & -0.210 & 2.92\pm0.41^\dagger & \rm WISE & \rm S20\\
\rm J1746-2856 & 0.126 & -0.233 & 1168.0 & 7.4\times10^{2} & 8.43\pm2.53 & N & 0.320 & -0.210 & 2.92\pm0.41^\dagger & \rm WISE & \rm S20\\
\rm J1753-2501 & 4.274 & 0.512 & 672.0 & 55 & 9.28\pm2.78 & N & 4.290 & 0.550 & 1.30\pm0.40 & \rm HH14 & \rm S22\\
\rm J1757-2421 & 5.281 & 0.054 & 179.5 & 0.26 & 4.40\pm1.32 & N & 5.332 & 0.081 & [2.89,13.40] & \rm WISE & \\
\rm J1801-2304 & 6.837 & -0.066 & 1067.8 & 3.9\times10^{2} & 12.61\pm3.78 & N & 6.979 & -0.250 & 2.70\pm0.80 & \rm HH14 & \\
\rm J1811-1736 & 12.821 & 0.435 & 473.9 & 46 & 5.99\pm1.80 & N & 12.762 & 0.370 & 2.40\pm0.20 & \rm HH14 & \rm S40\\
\rm J1812-1718 & 13.109 & 0.538 & 251.4 & 31 & 4.15\pm1.24 & N & 12.762 & 0.370 & 2.40\pm0.20 & \rm HH14 & \rm S40\\
\rm J1812-1733 & 12.904 & 0.387 & 509.8 & 59 & 6.32\pm1.89 & N & 12.762 & 0.370 & 2.40\pm0.20 & \rm HH14 & \rm S40\\
\rm J1813-1749 & 12.816 & -0.020 & 1087.0 & 4.1\times10^{3} & 12.05\pm3.62 & N & 12.807 & -0.204 & 2.40\pm0.17 & \rm HH14 & \rm W33\\
\rm J1816-1729 & 13.433 & -0.424 & 520.7 & 14 & 6.46\pm1.94 & N & 13.520 & -0.410 & 1.98\pm0.13^\dagger & \rm HH14 & \rm S43\\
\rm J1818-1422 & 16.405 & 0.610 & 619.6 & 64 & 7.33\pm2.20 & N & 16.936 & 0.758 & 2.00\pm0.25 & \rm HH14 & \rm S49\\
\rm J1820-1346 & 17.161 & 0.483 & 771.0 & 1.0\times10^{2} & 8.95\pm2.69 & N & 17.144 & 0.765 & 2.00\pm0.25 & \rm HH14 & \rm multiple\\
\rm J1822-1400 & 17.252 & -0.176 & 649.3 & 6.8 & 4.87\pm1.46 & N & 17.250 & -0.195 & >3.90 & \rm HH14 & \\
\rm J1823-1115 & 19.767 & 0.946 & 428.6 & 6.1 & 5.70\pm1.71 & N & 18.426 & 1.922 & [2.50,13.00] & \rm WISE & \\
\rm J1824-1159 & 19.253 & 0.324 & 463.0 & 18 & 5.67\pm1.70 & N & 19.485 & 0.138 & 2.00\pm0.20 & \rm HH14 & \\
\rm J1824-1423 & 17.146 & -0.796 & 427.6 & 12 & 5.70\pm1.71 & N & 16.808 & -1.072 & 1.98\pm0.13^\dagger & \rm HH14 & \rm S50\\
\rm J1825-1446 & 16.805 & -1.001 & 352.2 & 21 & 5.16\pm1.55 & N & 16.808 & -1.072 & 1.98\pm0.13^\dagger & \rm HH14 & \rm S50\\
\rm J1826-1131 & 19.800 & 0.293 & 320.6 & 1.1 & 4.64\pm1.39 & N & 18.725 & -0.045 & [4.30,10.90] & \rm WISE & \\
\rm J1832-1021 & 21.587 & -0.597 & 474.1 & 14 & 5.88\pm1.76 & N & 21.902 & -0.368 & 2.50\pm0.40 & \rm HH14 & \rm multiple\\
\rm J1833-0559 & 25.514 & 1.321 & 346.7 & 1.8\times10^{2} & 6.83\pm2.05 & Y & 25.647 & 1.054 & >3.00 & \rm HH14 & \\
\rm J1833-0827 & 23.386 & 0.063 & 410.9 & 1.1 & 4.71\pm1.41 & N & 23.115 & 0.556 & 1.70\pm0.50 & \rm HH14 & \\
\rm J1835-0643 & 25.093 & 0.552 & 464.8 & 1.5\times10^{2} & 6.18\pm1.85 & N & 24.919 & 0.294 & 5.70\pm0.60^\dagger & \rm WISE & \\
\rm J1837-0604 & 25.960 & 0.265 & 462.0 & 62 & 6.44\pm1.93 & N & 25.700 & 0.031 & 2.67\pm0.70 & \rm HH14 & \\
\rm J1841-0500 & 27.323 & -0.034 & 532.0 & 2.2\times10^{3} & 7.04\pm2.11 & N & 27.281 & -0.131 & 5.50\pm0.50 & \rm WISE & \\
\rm J1844-0030 & 31.711 & 1.271 & 603.2 & 12 & 10.41\pm3.12 & Y & 31.881 & 1.417 & 3.40\pm0.40 & \rm WISE & \\
\rm J1850-0006 & 32.764 & 0.093 & 655.0 & 2.6\times10^{2} & 8.56\pm2.57 & N & 32.823 & 0.072 & 7.10\pm0.10 & \rm WISE & \\
\rm J1850-0026 & 32.407 & 0.066 & 947.0 & 46 & 11.10\pm3.33 & N & 32.423 & 0.077 & [2.50,11.30] & \rm WISE & \\
\rm J1852+0031 & 33.523 & 0.017 & 787.0 & 3.6\times10^{2} & 9.58\pm2.87 & N & 33.507 & -0.002 & 8.80\pm3.00^\dagger & \rm WISE & \\
\rm J1852+0056g & 33.853 & 0.249 & 905.7 & 85 & 11.65\pm3.49 & N & 33.914 & 0.111 & <6.89 & \rm HH14 & \\
\rm J1852-0127 & 31.706 & -0.802 & 427.9 & 57 & 7.14\pm2.14 & N & 31.650 & -0.649 & 5.49\pm0.39^\dagger & \rm HH14 & \\
\rm J1853+0505 & 37.650 & 1.956 & 279.0 & 2.2\times10^{2} & 9.13\pm2.74 & Y & 37.642 & 1.192 & 1.88\pm0.08^\dagger & \rm WISE & \\
\rm J1853+0545 & 38.354 & 2.064 & 197.9 & 21 & 6.52\pm1.96 & Y & 38.123 & 1.660 & [2.29,10.60] & \rm WISE & \\
\rm J1854+0131g & 34.638 & -0.004 & 474.9 & 6.3\times10^{2} & 7.48\pm2.25 & N & 34.932 & -0.018 & 2.11\pm0.60 & \rm HH14 & \\
\rm J1855+0205 & 35.281 & 0.007 & 867.3 & 16 & 11.65\pm3.49 & N & 35.349 & 0.005 & [3.20,10.10] & \rm WISE & \\
\rm J1855+0422 & 37.314 & 1.052 & 455.6 & 54 & 10.95\pm3.28 & Y & 37.642 & 1.192 & 1.88\pm0.08^\dagger & \rm WISE & \\
\rm J1856+0113 & 34.560 & -0.497 & 96.1 & 1.6\times10^{-3} & 3.30\pm0.99 & Y & 34.757 & -0.669 & 2.99\pm0.40 & \rm WISE & \\
\rm J1856+0245 & 36.008 & 0.057 & 623.5 & 18 & 9.01\pm2.70 & N & 35.663 & -0.030 & 2.11\pm0.60 & \rm HH14 & \\
\rm J1856+0404 & 37.128 & 0.745 & 341.3 & 13 & 6.77\pm2.03 & N & 37.642 & 1.192 & 1.88\pm0.08^\dagger & \rm WISE & \\
\rm J1857+0210 & 35.586 & -0.393 & 783.0 & 27 & 10.94\pm3.28 & N & 35.588 & -0.489 & 2.11\pm0.60 & \rm HH14 & \\
\rm J1857+0212 & 35.617 & -0.390 & 506.8 & 7.9 & 8.00\pm2.40 & Y & 35.588 & -0.489 & 2.11\pm0.60 & \rm HH14 & \\
\rm J1857+0526 & 38.438 & 1.187 & 464.8 & 25 & 12.23\pm3.67 & Y & 37.642 & 1.192 & 1.88\pm0.08^\dagger & \rm WISE & \\
\rm J1858+0215 & 35.725 & -0.493 & 702.0 & 63 & 10.14\pm3.04 & N & 35.588 & -0.489 & 2.11\pm0.60 & \rm HH14 & \\
\rm J1858+0346 & 37.083 & 0.182 & 386.0 & 42 & 7.00\pm2.10 & N & 37.032 & 0.139 & - & \rm WISE & \rm multiple\\
\rm J1901+0156 & 35.818 & -1.367 & 105.4 & 2.1\times10^{-2} & 3.23\pm0.97 & Y & 35.673 & -0.847 & 2.11\pm0.60 & \rm HH14 & \\
\rm J1901+0331 & 37.213 & -0.637 & 402.1 & 1.3 & 7.29\pm2.19 & N & 37.370 & -0.367 & [2.56,10.40] & \rm WISE & \\
\rm J1901+0716 & 40.569 & 1.056 & 252.8 & 0.20 & 5.47\pm1.64 & N & 40.160 & 1.510 & >1.77 & \rm HH14 & \\
\rm J1903+0135 & 35.727 & -1.955 & 245.2 & 0.24 & 3.35\pm1.00 & N & 36.289 & -1.686 & 1.80\pm0.50 & \rm HH14 & \\
\rm J1907+0740 & 41.613 & -0.102 & 332.0 & 0.24 & 6.85\pm2.06 & N & 41.379 & 0.037 & >4.12 & \rm HH14 & \\
\rm J1910+0534 & 40.056 & -1.668 & 484.0 & 25 & 21.25\pm6.38 & Y & 39.904 & -1.331 & 2.10\pm0.60 & \rm HH14 & \rm multiple\\
\rm J1913+1000 & 44.285 & -0.194 & 422.0 & 22 & 7.86\pm2.36 & N & 44.379 & -0.326 & 6.10\pm1.30 & \rm WISE & \\
\rm J1916+1030 & 45.099 & -0.638 & 387.2 & 18 & 8.64\pm2.59 & N & 45.002 & -0.610 & 6.00\pm1.10 & \rm WISE & \\
\rm J1924+1631 & 51.405 & 0.318 & 518.5 & 49 & 13.98\pm4.19 & N & 51.010 & 0.060 & 7.20\pm1.20 & \rm WISE & \\
\rm J1946+2535 & 61.809 & 0.283 & 248.8 & 4.5 & 8.30\pm2.49 & Y & 61.467 & 0.380 & 3.90\pm0.10 & \rm WISE & \\
\rm J2005+3411g & 71.274 & 1.239 & 489.0 & 43 & 22.75\pm6.83 & N & 71.312 & 0.828 & >2.66 & \rm HH14 & \\
\rm J2013+3845 & 75.930 & 2.476 & 238.2 & 0.13 & 8.43\pm2.53 & N & 75.460 & 2.430 & 1.20\pm0.50 & \rm HH14 & \\
\rm J2021+4024g & 78.157 & 2.107 & 680.5 & 74 & 25.00\pm7.50 & Y & 78.698 & 1.902 & 1.50\pm0.10^\dagger & \rm WISE & \\
\rm J2022+3842 & 76.888 & 0.960 & 429.1 & 55 & 22.06\pm6.62 & N & 77.402 & 0.841 & 1.50\pm0.10^\dagger & \rm WISE & \\
\rm J2022+3845g & 76.906 & 1.015 & 487.5 & 1.8\times10^{2} & 21.48\pm6.44 & N & 77.402 & 0.841 & 1.50\pm0.10^\dagger & \rm WISE & \\
\rm J2029+3744 & 76.898 & -0.727 & 190.7 & 4.9\times10^{-2} & 6.33\pm1.90 & N & 77.402 & 0.841 & 1.50\pm0.10^\dagger & \rm WISE & \\
\rm J2030+3944g & 78.625 & 0.294 & 937.4 & 1.0\times10^{2} & 25.00\pm7.50 & Y & 78.689 & 0.355 & 1.50\pm0.10^\dagger & \rm WISE &\\
\rm J2052+4421g & 84.833 & -0.168 & 547.0 & 81 & 25.00\pm7.50 & Y & 85.325 & -0.788 & 1.50\pm0.10^\dagger & \rm WISE & \rm S117\\
\rm J2108+4441 & 86.909 & -2.012 & 139.8 & 0.11 & 4.96\pm1.49 & N & 85.325 & -0.788 & 1.50\pm0.10^\dagger & \rm WISE & \rm S117\\
\enddata
\tablecomments{Pulsar name, Galactic coordinates, DMs, scattering time ($\tau$) in ms at 1 GHz, distance ($D$), and the method used to derive the pulsar distance (P for parallax, Y for YMW16, and N for NE2001). For YMW16 and NE2001, $30\%$ distance errors are shown. For \HII regions we show Galactic coordinates, distances, and the corresponding catalog reference (HH14 for \citealt{houhan2014} and WISE for \citealt{wisecat_2014}). \HII regions with parallax distances are indicated by a $\dagger$. Distance lower limits correspond to cases where HH14 only reports a near kinematic distance but no kinematic distance ambiguity resolution. Kinematic distances without an ambiguity resolution (e.g. from HI or H$_2$CO) are shown as a range spanning the near to far distances. In these cases we show intersections for pulsars with estimated distances greater than the near distance, although we note that these intersections should be treated with caution. The Notes column indicates \HII region names where possible, and pulsars for which multiple \HII regions may be relevant are noted as ``multiple.''}
\end{deluxetable*}

\clearpage

\onecolumngrid
\section{Pulsars with Excess Scattering}\label{app:outliers}

We have identified $\sim 30$ pulsars with significant scattering excesses and deficits from the mean $\tau$-DM relation for the pulsar population (\S\ref{sec:candidates}). Table~\ref{tab:outliers} lists these outliers, their properties, and the foreground structures that may be related to their scattering excesses and deficits.

\begin{deluxetable*}{l C C C C c}[htb!]\label{tab:outliers}
\tabletypesize{\footnotesize}
\tablecaption{Pulsars with Significant Scattering Excess or Deficit}
\tablehead{\colhead{PSR} & \colhead{$l$ ($^\circ$)} & \colhead{$b$ ($^\circ$)} & \colhead{DM (pc cm$^{-3}$)} & \colhead{$\tau$ (ms at 1 GHz)} & \colhead{Candidate Foreground Structure}}
\startdata
\cutinhead{$\tau > 2\sigma$ from Mean $\tau$-DM Relation}
J0450$-$1248 & 211.075 & -32.629 & 37.0 & 0.029 & S276; maybe Orion-Eridanus superbubble \\
J0502$+$4654 & 160.363 & 3.077 & 41.8 & 0.085 & Supernova Remnant (G160.9$+$02.6)$^a$ \\
J0540$-$6919 & 279.717 & -31.516 & 147.2 & 1.8 & Supernova Remnant (SNR 0540$-$693)$^b$ \\
J0646$+$0905 & 204.271 & 3.051 & 147.9 & 10.0 & \HII Region (S273) \\
J0835$-$4510 & 263.552 & -2.787 & 67.8 & 0.049 & Gum Nebula$^c$ \\
J1514$-$5925$^\dagger$ & 320.284 & -1.482 & 192.5 & 16 & Multiple \HII Regions \\
J1550$-$5418$^\dagger$ & 327.237 & -0.132 & 830.0 & 4.3\times10^3 & Multiple \HII Regions \\
J1643$-$1224$^\dagger$ & 5.669 & 21.218 & 62.4 & 4.9\times10^{-2} & \HII Region (S27)$^d$ \\
J1730$-$3350 & 354.133 & 0.090 & 261.3 & 27.3 & \HII Region (G354.175$-$00.062)\\
J1812$-$1718$^\dagger$ & 13.109 & 0.538 & 251.4 & 31 & \HII Region (S40) \\
J1819$-$1114 & 19.291 & 1.857 & 309.7 & 51 & \HII Region (S54) \\
J1833$-$0559$^\dagger$ & 25.514 & 1.321 & 346.7 & 1.8\times10^{2} & \HII Region (IRAS 18316$-$0602) \\
J1834$-$0731 & 24.288 & 0.366 & 288.3 & 122 & Multiple \HII Regions \\
J1841$-$0500$^\dagger$ & 27.323 & -0.034 & 532.0 & 2.2\times10^{3} & \HII Region (G027.281$-$0.132) \\
J1849$-$0014g & 32.505 & 0.321 & 346.6 & 73 & \nodata \\
J1849$+$0127 & 34.034 & 1.043 & 207.0 & 9.7 & \nodata \\
J1853$+$0505$^\dagger$ & 37.650 & 1.956 & 279.0 & 2.2\times10^{2} & \HII Region (G037.642$+$1.192)\\
J1853$+$0545$^\dagger$  & 38.354 & 2.064 & 197.9 & 21  & \HII Region (G038.123$+$1.660) \\
J1854$+$0131g$^\dagger$ & 34.638 & -0.004 & 474.9 & 6.3\times10^{2} & \HII Region (G034.932$-$0.018)\\
J1859$+$0601 & 39.245 & 0.903 & 272.4 & 125 & \HII Region (G039.283$+$00.865) \\
J1905$+$0616 & 40.069 & -0.169 & 256.1 & 27 & \HII Region (G039.294$-$00.311) \\
J1919$+$1745 & 51.897 & 1.987 & 142.3 & 2.9 & \nodata \\
J1920$+$1110 & 46.152 & -1.199 & 188.4 & 25 & \nodata \\
J1921$+$1419 & 49.058 & 0.021 & 91.6 & 8 & \HII Region (W51) \\
J1949$+$2306 & 59.931 & -1.423 & 196.3 & 5.4 & \HII Region (S86) \\
J1953$+$2732 & 64.205 & 0.059 & 194.2 & 21 & \nodata \\ 
\cutinhead{$\tau < 2\sigma$ from Mean $\tau$-DM Relation}
J0953$+$0755 & 228.908 & 43.697 & 2.9 & 1.1\times10^{-8} & Local Bubble$^e$ \\
J1017$-$5621 & 282.732 & 0.341 & 438.7 & 0.11 & \nodata \\
J1709$-$4429 & 343.098 & -2.686 & 75.6 & 3.6\times10^{-5} & Pulsar Wind Nebula (G343.1$-$2.3)$^f$ \\
J1932$+$2220 & 57.356 & 1.554 & 218.9 & 0.006 & \nodata \\
\enddata
\tablecomments{Pulsars that have $\tau$ more than $2\sigma$ deviant from the mean $\tau$-DM relation (\S\ref{sec:candidates}), and the candidate structures that may explain their scattering excesses and deficits. Pulsars positively identified as intersecting \HII regions by our selection criteria are indicated by a $\dagger$. Scattering deficits could be explained by underdense voids along the LOS or dense structures close enough to the pulsar that they contribute excess DM but no significant scattering. {References: (a) \cite{jing2023}; (b) \cite{seward1984}; (c) \cite{mitra2001}; (d) \cite{mall2022}; (e) \cite{bhat1998}; (f) \cite{romani2005}.}}
\end{deluxetable*}

\end{document}